\documentclass[numberedheadings]{aip-cp}

\usepackage[numbers,sort&compress]{natbib}
\usepackage[table]{xcolor}
\usepackage{url}
\usepackage{fancyhdr}
\usepackage{rotating}
\usepackage{graphicx}
\usepackage{comment}

\usepackage{setspace}
\usepackage{fancyhdr}
\fancypagestyle{plain}{
    \fancyhf{} 
     
    \fancyfoot[R]{\thepage} 
}
\begin{document}

\title{Can G4-like Composite \textit{Ab Initio} Methods Accurately Predict Vibrational Harmonic Frequencies?}

\author{Emmanouil Semidalas}
\author{Jan M. L. Martin}
\eaddress{gershom@weizmann.ac.il}

\affil{$^a$ Dept. of Molecular Chemistry and Materials Science, Weizmann Institute of Science, 7610001 Re\d{h}ovot, Israel.}

\maketitle

\begin{abstract}
Minimally empirical G4-like composite wavefunction theories [E. Semidalas and J. M. L. Martin, \textit{J. Chem. Theory Comput.} {\bf 16}, 4238–4255 and 7507-7524 (2020)] trained against the large and chemically diverse GMTKN55 benchmark suite have demonstrated both accuracy and cost-effectiveness in predicting thermochemistry, barrier heights, and noncovalent interaction energies. Here, we assess the spectroscopic accuracy of top-performing methods: G4-\textit{n}, cc-G4-\textit{n}, and G4-\textit{n}-F12, and validate them against explicitly correlated coupled-cluster CCSD(T*)(F12*) harmonic vibrational frequencies and experimental data from the HFREQ2014 dataset, of small first- and second-row polyatomics. G4-T is three times more accurate than plain CCSD(T)/def2-TZVP, while G4-T$_{\rm ano}$ is two times superior to CCSD(T)/ano-pVTZ. Combining CCSD(T)/ano-pVTZ with MP2-F12 in a parameter-free composite scheme results to a root-mean-square deviation of ~5 cm$^{-1}$ relative to experiment, comparable to CCSD(T) at the complete basis set limit. Application to the harmonic frequencies of benzene reveals a significant advantage of composites with ANO basis sets -- MP2/ano-pV\textit{m}Z and [CCSD(T)-MP2]/ano-pVTZ (\textit{m} = Q or 5) -- over similar protocols based on CCSD(T)/def2-TZVP. 
Overall, G4-type composite energy schemes, particularly when combined with ANO basis sets in CCSD(T), are accurate and comparatively inexpensive tools for computational vibrational spectroscopy. \\
\\
{\bf Keywords:} coupled-cluster theory, vibrational spectroscopy, composite wavefunction methods
\end{abstract}

\section{Introduction}
In the realm of computational spectroscopy,\cite{Yurchenko2023} accurate prediction of molecular vibrations has long been a valuable tool in chemistry, biochemistry, and materials science.\cite{Puzzarini2010,Puzzarini2019,Barone2021,Baiz2020,Steinhauser2009,Paul2019} To achieve spectroscopic accuracy, defined as an error of less than 1 cm$^{-1}$ from gas-phase vibrations,\cite{Stanton2003,jmlm230} it may be necessary to venture beyond the ‘gold-standard’ CCSD(T) method,\cite{Raghavachari1989} i.e. coupled-cluster with single, double, and perturbative triple excitations, and employ higher-order correlation approaches that include quadruple excitations, such as in the CCSDT(Q) method.\cite{Kallay2005} Additionally, the slower basis set convergence\cite{Schwartz1962a,Hill1985,Kutzelnigg1992a} of the correlation energy can quickly lead to computational constraints, especially for the electronic structure of larger molecules.

In response to the challenges that lie with computationally expensive \textit{ab initio} methods, the composite wave function theories (cWFTs) have emerged as promising alternatives (for a very recent review, see Karton\cite{Karton2022}). These methods combine high-level treatments of electron correlation through additivity approximations, economical basis set extrapolations, and often empirical corrections. The end result is a robust approach aimed at the accuracy of CCSD(T) at the complete basis set limit (CBS). Examples of such cWFTs include the Pople's group Gaussian-n theories (G\textit{n}),\cite{Pople1989,Curtiss1991,Curtiss1995,Baboul1999,Curtiss2007,Curtiss2007a} the CBS models by Petersson and co-workers,\cite{Montgomery1999,Montgomery2000,Petersson2001} 
the Weizmann-\textit{n} theory (\textit{n}= 1 and 2) by the Martin group,\cite{Martin1999,jmlm148,jmlm151} and the ccCA (correlation consistent composite approach) of Wilson and coworkers.\cite{DeYonker2006,DeYonker2009,Peterson2016}

Among the more economical composite energy schemes, G4\cite{Curtiss2007a} and G4(MP2)\cite{Curtiss1993,Curtiss1999,Curtiss2007} approaches fit in. The former combines MP\textit{n} (\textit{n} = 2 and 4) and CCSD(T) methods, while the latter reduces cost by omitting the MP4 step. In 2011, Radom and co-workers introduced the G4(MP2)-6X protocol,\cite{Chan2011} an improved G4(MP2) variant, featuring six empirical parameters for correlation energies and another six for the high-level correction. Building on this, Chan et al.\cite{Chan2019b} shifted from Pople- to Karlsruhe-type basis sets in their G4(MP2)-XK approach. Inspired by their work, we presented an hierarchy of G4-type cWFTs\cite{Semidalas2020a,Semidalas2020}, validated against the energetics of the chemically large GMTKN55 benchmark suite.\cite{Goerigk2017a} (general main-group thermochemistry, kinetics, and noncovalent interactions, 55 subsets)

Another avenue for improving cost-effectiveness and accuracy in cWFTs is using explicitly correlated theory,\cite{Ten-no2012,Ten-no2012a,Kong2012,Hattig2012a} where \textit{r}$_{12}$ terms that depend on the interelectronic distance, e.g. of the form [1-exp($\gamma$/r$_{12}$)]/$\gamma$, are added in the wave function. This inclusion \mbox{accelerates} basis set convergence,\cite{Klopper1987,Kutzelnigg1991} with R12/F12 methods typically requiring 2-3 additional basis set \mbox{cardinal} numbers or "zetas" compared to conventional calculations.\cite{Hattig2012a,Kong2012,Ten-no2012} Presently, numerous \textit{explicitly correlated} \mbox{composite} thermochemical protocols have been reported, including the W4-F12,\cite{Karton2012} ccCA-F12,\cite{Mahler2013} G4-\textit{m}-F12,\cite{Semidalas2020} and SVECV-f12 theories.\cite{Ventura2021}

For harmonic vibrational frequencies it is generally accepted that valence-only CCSD(T) suffices, particularly in systems with low static correlation that are dominated by a single reference determinant.\cite{Stanton1997,Stanton2003} This arises from a fortuitous error compensation between the approximate treatment of the triples term, the missing core-valence correlation, and the neglect of higher order excitations.

To approach the full CI limit with composite energy schemes, post-CCSD(T) terms must be included. This has been successfully demonstrated in the well-known HEAT-n approaches,\cite{Tajti2004,Bomble2006,Harding2008,Thorpe2019} the Feller-Peterson-Dixon (FPD) model,\cite{Dixon2012,Feller2014,Feller2016} the Wilson's group ccCA,\cite{DeYonker2006,DeYonker2009,Peterson2016} and the Weizmann-n (n = 3 and 4) theories\cite{jmlm173,Karton2006,Sylvetsky2016} by the Martin group. 
We recently examined\cite{Spiegel2023} the importance of post-CCSD(T) corrections in cWFTs, particularly CCSDT(Q)$_{\Lambda}$,\cite{Kallay2005} for spectroscopic constants in heavy-atom  diatomics at different static correlation regimes, and reported accurate predictions, including ozone vibrational frequencies.

There has been a fair amount of work on post-CCSD(T) cWFT methods in the context of vibrational spectroscopy. We note, inter alia, the 2005 work of Heckert et al.\cite{Heckert2005,Heckert2006} and Puzzarini et al.\cite{Puzzarini2008} on accurate geometries {\em viz.} rotational constants.
Ruden et al.\cite{Ruden2004} considered quadruples and quintuples terms in CCSD(T)-based composite schemes for harmonic frequencies of HF, N$_2$, F$_2$, and CO, while Karton and Martin\cite{jmlm230} applied pointwise W4 theory (and truncations thereof, as well as the enhanced W4.3 theory) to spectroscopic constants and electric properties of 28 first- and second-row diatomics, as well as several polyatomics.\cite{jmlm230}
The spectroscopic constants of formaldehyde were obtained by Schaefer and co-workers\cite{Morgan2018} through CCSDT(Q)-based focal-point analysis,\cite{Allen1993,East1993,East1993a,Csaszar1998} while Zhu and Xu\cite{Zhu2023} reported static polarizabilities at CCSD(T)/CBS. Huang and Lee,\cite{Huang2008}, and later Lee and coworkers\cite{Fortenberry2019,Gardner2021} explored the CcCR methodologies ("C" stands for CBS, complete basis set; "cC" for core correlation; and "R" for relativistic effects) for determining fundamental vibrational frequencies\cite{Fortenberry2019} and anharmonic rotational constants.\cite{Gardner2021} 

In contrast, as well as in comparison to combining high-quality harmonic frequencies with DFT-level anharmonic force fields (see, e.g.,\cite{Barone2021} and references therein; Ref.\cite{jmlm172}), effort toward an economical cWFT approximation to CCSD(T) harmonic frequencies has been fairly limited. An unfairly overlooked paper by Bettens and coworkers\cite{Bettens2006} considered the combination of MP2 in larger basis sets with CCSD(T)$-$MP2 in smaller ones. Barone and coworkers\cite{Barone2014} introduced what they termed their `cheap' approximation, which augments an MP2/cc-pV\{T,Q\}Z CBS extrapolation (the notation means `from cc-pVTZ and cc-pVQZ') with diffuse function, CCSD(T)$-$MP2, and core-valence corrections.

The purpose of this study is to assess whether G4-type composite energy schemes can be a viable alternative to large basis set CCSD(T) harmonic vibrational frequency calculations. 
We shall validate these G4-type methods against basis set limit extrapolated CCSD(T) frequencies, as well as 
CCSD(T*)(F12*)/VQZ-F12 calculated harmonic vibrational frequencies and experimental ones of 31 molecules from the HFREQ2014 dataset.\cite{Martin2014}  
As a proof of principle, cWFTs are then applied to the difficult\cite{MTLbenzene} harmonic force field of benzene.

\section{Computational details}
All calculations were performed on the Chemfarm HPC cluster of the Faculty of Chemistry at the Weizmann Institute, mostly using the MOLPRO 2022.3\cite{Werner2020} electronic structure program system.
Built on top of the ALASKA integral derivative package,\cite{Lindh1993}
canonical MP2\cite{Moeller1934} analytical derivatives (Ref.\cite{HeadGordon1999} and references therein; see Ref.\cite{ElAzhary1998} for the specific MOLPRO implementation) and canonical CCSD(T)\cite{Raghavachari1989, Watts1993} analytical first derivatives (Ref.\cite{Harding2008} and references therein) were evaluated with nondegenerate symmetry enabled, while 
force constant matrices (Hessians) were evaluated semi-numerically using central differences of gradients.
For verification purposes, MP2 Hessians in the same basis set were also calculated analytically\cite{HeadGordon1994} using Gaussian 16;\cite{gaussian16} we found harmonic frequencies from the analytical and semi-numerical Hessians to differ by only on the order of 0.03 cm$^{-1}$, which is negligible in the context of this work.

The explicitly correlated density-fitted DF-MP2-F12 method\cite{Womack2014} was employed with analytic gradients\cite{Gyorffy2017,Gyorffy2018} and the 3*C(FIX, HY1) \textit{Ansatz}, in which the extended Brillouin condition is assumed and the "HY1" hybrid approximation is used for matrix elements\cite{Werner2007} over the F12 geminal,\cite{Ten-no2004} together with fixed geminal amplitudes.\cite{Ten-no2004,Hill2009a} The CCSD(T)(F12*)\cite{Hattig2010} geometry optimizations and frequency calculations were carried out fully numerically for want of an analytical gradient.

In our study, we employed various basis set families, including the Weigend-Ahlrichs def2 family\cite{Weigend2005}: def2-SVP, def2-nZVP, and def2-nZVPP,  (n = T and Q), along with their augmented alternatives with diffuse functions def2-SVPD and def2-nZVPPD (n = T and Q).\cite{Rappoport2010} The combination of def2-nZVPP on hydrogen atoms and def2-nZVPPD on main group elements is denoted as def2-nZVPPD’. Among the atomic natural orbital (ANOs) basis sets pioneered by Alml\"of and Taylor,\cite{Almlof1987} we chose the ano-pVnZ (n = D,T,Q,5) of Neese  and Valeev,\cite{Neese2011} as well as their aug-ano-pVnZ (diffuse function augmented) and saug-ano-pVnZ (minimally augmented) variants from the same reference. Table \ref{tab:abbrev} provides a list of abbreviations for methods and basis sets used in this study. 

Among the correlation consistent basis set family, we consider the cc-pVnZ and aug-cc-pVnZ basis sets (n=D,T,Q,5)\cite{DunningJr1989,Kendall1992,Woon1993} on hydrogen and the first row, and the (aug-)-cc-pV(n+d)Z basis sets\cite{Dunning2001} on second-row elements, which include an additional tight $d$ function as was previously found\cite{jmlm107,jmlm191} to be important when these elements are in high oxidation states. (In this work, the largest impact is seen for SO$_2$.) Additionally, for calculations including inner-shell correlation, we employed the core-valence weighted aug-cc-pwCVnZ (n = T and Q) basis sets.\cite{Peterson2002} The shorthand haVnZ+d refers in this paper to the combination of cc-pVnZ on hydrogen with aug-cc-pVnZ on first-row atoms and aug-cc-pV(n+d)Z on second-row atoms.

Aside from the orbital basis set (OBS) employed in a standard explicitly correlated calculation with density-fitting, there are three additional auxiliary basis sets (ABS): the ‘JKFit’ basis set for the Coulomb and exchange integrals, the ‘MP2Fit’ basis set for density fitting in MP2, and the 'CABS' also known as complementary auxiliary basis set.\cite{Klopper2002a,Valeev2004} We utilized the cc-pVnZ-F12\cite{Peterson2008}  (n = T and Q) basis sets as OBS, along with the default cc-pVnZ-F12/JKFit and cc-pVnZ-F12/MP2Fit in MOLPRO as JKFit and MP2Fit ABSs, respectively. For CABS, we used Yousaf and Peterson's cc-pVnZ-F12/OptRI.\cite{Yousaf2009} Slater-type geminal terms of the F12 form [1-exp ($\gamma$ r$_{12}$)]/$\gamma$ were used with a $\beta$ geminal exponent of 1.0 for both triple- and quadruple-$\zeta$ OBS, as recommended in Table V of Ref.\cite{Hill2009a} In the text below, "VnZ-F12" signifies the cc-pVnZ-F12 basis sets.

\begin{table}[h!]
\centering
\label{tab:abbrev}
\begin{tabular}{lp{10cm}c}
\hline
Acronym & Meaning & Reference \\
\hline
MP2 & second-order M{\o}ller-Plesset perturbation theory (PT) & \cite{Moeller1934}\\
MP2-F12 & explicitly correlated second-order M{\o}ller-Plesset PT & \cite{Womack2014}\\
VPT2 & second-order vibrational PT & \cite{Nielsen1951}\\
CCSD(T) & coupled-cluster with single, double, and \\
        & quasiperturbative triple excitations & \cite{Raghavachari1989, Watts1993}\\
CCSD(T*)(F12*) & ditto but explicitly correlated coupled-cluster theory with \\
& the F12 geminal in the H{\"a}ttig-Kohn-Tew (F12*) approximation \\
& to CCSD-F12 and the Marchetti-Werner scaling for (T) & \cite{Hattig2010,Marchetti2009}\\
cWFT & composite wavefunction theory & \\
G\textit{n} & Gaussian-\textit{n} theory & \cite{Pople1989,Curtiss1991,Curtiss1995,Curtiss1998a,Baboul1999,Curtiss2007a,Curtiss2007}\\
G4-\textit{n} & G4-like theory with CCSD(T) contributions from \\
 & def2-SVPD or def2-TZVP basis sets (n=D,T) & \cite{Semidalas2020a}\\
cc-G4-\textit{n} & G4-like theory with core-valence correlation\\
& at the MP2 Level (n=D,T) & \cite{Semidalas2020}\\
G4-\textit{n}-F12 & explicitly correlated RI-MP2-F12-based G4-like theory & \cite{Semidalas2020}\\
def2-\textit{n}ZVPP & Weigend-Ahlrichs def2 basis set family (n=T,Q) & \cite{Weigend2005}\\
cc-pV\textit{n}Z & correlation-consistent basis set family for \\
 & valence correlation (n=D,T,Q)& \cite{DunningJr1989,Kendall1992,Woon1993}\\
cc-pV\textit{n}Z-F12 & correlation-consistent basis set family for \\
        & use in explicitly correlated calculations (n=D,T,Q) & \cite{Peterson2008}\\
ano-pV\textit{n}Z & atomic natural orbital basis sets (n=D,T,Q,5) & \cite{Neese2011}\\
GMTKN55 & general main-group thermochemistry, kinetics, &\\
&and noncovalent interactions, 55 subsets & \cite{Goerigk2017a}\\
WTMAD2 & weighted mean absolute deviation (type 2) & \cite{Goerigk2017a}\\
\hline
\end{tabular}
\caption{Abbreviations for methods, basis sets, and other terms in the present work}
\end{table}

To validate the accuracy of composite schemes, we used the HFREQ2014 dataset\cite{Martin2014} of harmonic frequencies for small molecules (Table \ref{tab:hfreq2014molecules}). Error statistics were estimated relative to CCSD(T*)(F12*)/VQZ-F12 calculations (reference) and experimental values from ref. \cite{Martin2014} and references therein. On a related note, Mehta et al.\cite{Mehta2023} considered the same CCSD(T*)(F12*)/VQZ-F12 reference for HFREQ2014 in their study on the performance of double-hybrid density functional theory for molecular vibrations; they also carried out CCSD(T) calculations there for comparison, but excluded some of the HFREQ2014 species such as F$_2$, HNO, and CF$_2$. Also, Head-Gordon and coworkers\cite{Liang2023} recently introduced analytical second derivatives of VV10 dispersion corrected\cite{Vydrov2010} containing density functionals and evaluated their predictive accuracy for harmonic frequencies across various molecular systems including those in the HFREQ2014 dataset. They concluded that while the VV10-enhanced DFT functionals offered no advantage for small-molecule vibrational spectra, but a significant improvement was seen in vibrational spectra of noncovalent complexes.

\begin{table}[h!]
  \caption{Molecules considered in the HFREQ2014 dataset}
  \label{tab:hfreq2014molecules}
  \begin{tabular}{l}
  \hline
BH$_{3}$, C$_{2}$H$_{2}$, C$_{2}$H$_{4}$, CCl$_{2}$, CF$_{2}$, CH$_{3}$OH, CH$_{4}$, Cl$_{2}$, ClCN, \\
ClF, CO, CO$_{2}$, CS, CS$_{2}$, F$_{2}$, H$_{2}$CO, H$_{2}$CS, H$_{2}$O,\\
H$_{2}$S, HCl, HCN, HF, HNO, HOCl, N$_{2}$, NH$_{3}$, NNO,\\
OCS, PH$_{3}$, SiO, SO$_{2}$ \\
\hline
\end{tabular}
\end{table}

Geometries were optimized using the total electronic energy as the target function for each cWFT method, employing the numerical gradient. That was accomplished through the optg procedure\cite{Eckert1997} in MOLPRO. The optimizations were completed once the maximum gradient component was less than 10$^{-5}$ hartree/bohr, the optimization step was less than 10$^{-5}$ bohr, and the change in total energy from the previous iteration was less than 10$^{-11}$ hartree.  In numerical gradients and Hessians, the default stepsize of 0.01 a.u. was used, unless otherwise noted (see Table \ref{tab:rmsdshfreqccsdt} below for cases where a 0.005 a.u. value was used). The cutoffs of two-electron integrals were set to 10$^{-20}$ for screening and 10$^{-18}$ for the prefactor test. The total energy in subsequent calculations of force constant calculations was converged within 10$^{-12}$ E$_h$.

\begin{figure}[h]
\label{fig:hfreq-diagram}
\caption{\texttt{hfreq} implementation workflow}
\includegraphics[width=2.1in]{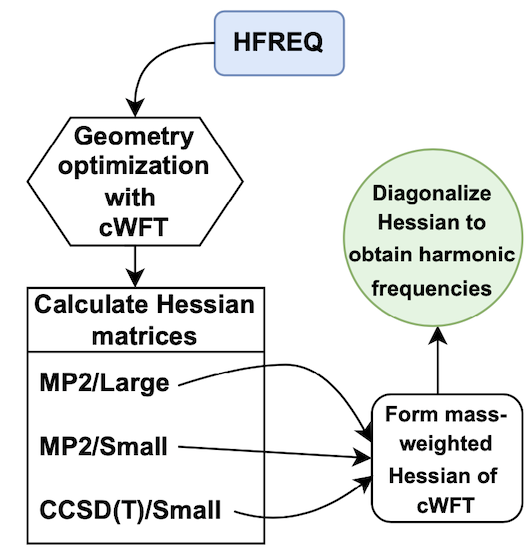}
\end{figure}

We provide an implementation of \texttt{hfreq} for automated geometry optimization and calculation of harmonic vibrational frequencies with composite energy schemes at the following Github link 'https://github.com/msemidalas/hfreq'. \texttt{hfreq} is written in Python and the Git repository contains several sample files to reproduce the results of this work. In this version, geometry optimization employs analytic gradients, while Hessians are computed numerically in MOLPRO. The mass-weighted Hessian is diagonalized in Psi4.\cite{Smith2020} Figure \ref{fig:hfreq-diagram} graphically shows the automated procedure for the evaluation of harmonic frequencies. 

A reviewer highlighted a very recent paper by Jensen,\cite{Jensen2023} in which three methods for extrapolating vibrational frequencies are discussed. The first, 'Opt-xpol,' involves geometry optimization by minimizing the basis set-extrapolated energy, followed by frequency calculation from the extrapolated Hessian at the optimized geometry: this parallels our approach in the present work and in the hfreq code. The second, 'v-xpol,' directly extrapolates vibrational frequencies from optimized geometries using two different basis sets, an approach explored earlier by Varandas\cite{Varandas2022} and Broda and colleagues.\cite{Buczek2011a,Buczek2011b,Broda2012} The 'H-xpol' approach directly extrapolates optimized Hessians, regardless of reference geometries. Jensen's findings show that all three approaches yield similar results for small molecules using double-triple $\zeta$ extrapolation in cc-pVnZ basis sets at wB97X-D\cite{Chai2008a} and MP2 levels. However, for H-bonded complexes, 'H-xpol' yields unsatisfactory results with extrapolation from pcseg-0 and pcseg-1 basis sets,\cite{Jensen2001} as well as from pcseg-1 and pcseg-2, likely due to poor reference geometries.

\section{Results and discussion}
\subsection{Performance of CCSD(T) for harmonic frequencies}
Assessing CCSD(T) for harmonic frequencies is the first step to estimate the accuracy of composite energy schemes. Kesharwani and Martin reported that valence-only CCSD(T) at the complete basis set limit is about  5 cm$^{-1}$ as accurate as the experimental harmonic frequencies for the HFREQ2014 dataset.\cite{Martin2014} (For the avoidance of doubt, we should stress that these experimental data are truly harmonic, obtained from fitting series expansion in the vibrational quantum numbers to many vibrational band origins.) Explicitly-correlated CCSD(T*)(F12*) with VQZ-F12 achieves a root-mean-square deviation (RMSD) of 4.7 cm$^{-1}$ compared to experiment; T* denotes pointwise Marchetti-Werner scaling\cite{Marchetti2009} of the triples. In a recent DFT study of harmonic frequencies,\cite{Mehta2023} very similar results were obtained with fairly large basis sets, such as VQZ-F12, between either the point-wise scaling T* or scaling the triples term by a constant factor (T$_s$)\cite{jmlm261}.

\begin{figure}[htpb]
\label{fig:hfreq_CCSDt}
\caption{Box-and-whisker plot for the deviations of harmonic vibrational frequencies at CCSD(T) with various basis sets from CCSD(T*)(F12*)/VQZ-F12 for the HFREQ2014 dataset. The interquartile range (IQR) is the difference between the third and first quartiles (IQR = Q$_3$ – Q$_1$). The upper whisker extends up to Q$_3$ + 1.5*IQR, while the lower whisker extends down to Q$_1$ - 1.5*IQR, and outliers are shown outside the whiskers.  The median is indicated by a red line, while a green dotted line represents the mean. The blue band indicates $\pm10$ cm$^{-1}$.}
\includegraphics[width=5in]{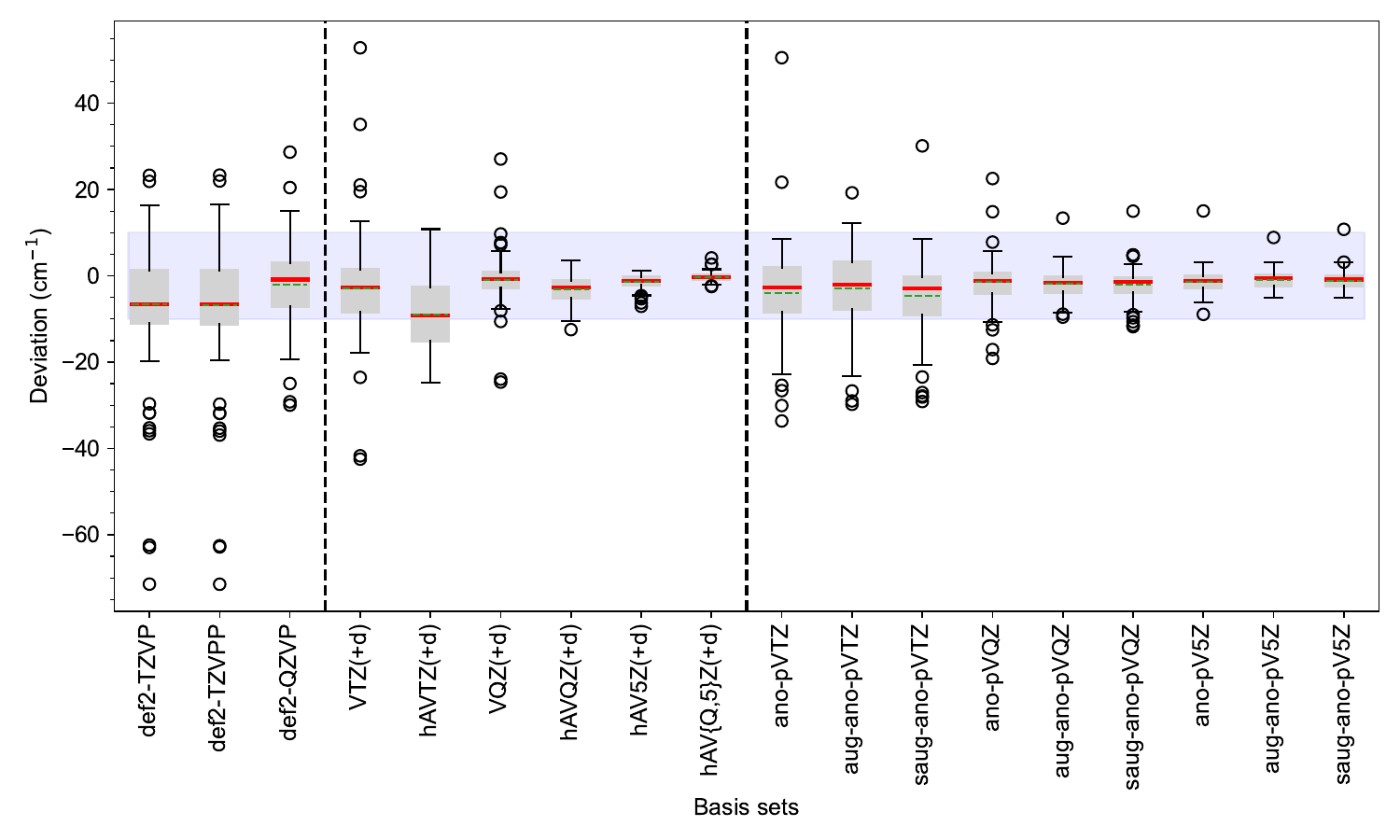}
\end{figure}

Table \ref{tab:rmsdshfreqccsdt} presents the error statistics of CCSD(T) with different classes of basis sets compared to reference calculated harmonic frequencies and experimental values for HFREQ2014; the error distribution is also depicted as a `box and whiskers plot' in Figure~\ref{fig:hfreq_CCSDt}. 

\begin{table}[]
  \centering
  \caption{Root-mean-square deviations and mean-absolute deviations (cm$^{-1}$) of calculated harmonic frequencies at the CCSD(T) level from CCSD(T*)(F12*)/VQZ-F12 calculations (refered to as CCSD(T)/CBS) and experiment for the HFREQ2014 dataset}
  \label{tab:rmsdshfreqccsdt}
   \footnotesize
    \begin{tabular}{lp{0.6cm}p{0.6cm}p{0.6cm}p{0.6cm}|lp{0.6cm}p{0.6cm}p{0.6cm}p{0.6cm}|lp{0.6cm}p{0.6cm}p{0.6cm}p{0.6cm}}
    \hline
    & \multicolumn{4}{c|}{Errors relative to} & & \multicolumn{4}{c|}{Errors relative to} & & \multicolumn{4}{c}{Errors relative to}\\
    & \multicolumn{2}{c}{CCSD(T)/CBS} & \multicolumn{2}{c|}{Expt} & & \multicolumn{2}{c}{CCSD(T)/CBS} & \multicolumn{2}{c|}{Expt} & & \multicolumn{2}{c}{CCSD(T)/CBS} & \multicolumn{2}{c}{Expt}\\ 
    \hline
    CCSD(T) & RMSD & MAD & RMSD & MAD & CCSD(T) & RMSD & MAD & RMSD & MAD & CCSD(T) & RMSD & MAD & RMSD & MAD \\
    \hline
        def2-SVP & 31.06 & 23.82 & 30.36 & 23.22 & VDZ+d & 32.65 & 20.68 & 31.31 & 19.82 & ano-pVDZ & 28.71$^a$ & 18.89 & 26.99$^a$ & 17.41 \\
    def2-SVPD & 28.13 & 18.51 & 26.53 & 17.96 & haVDZ+d  & 37.93$^b$ & 27.63 & 36.72$^b$ & 27.13 & saug-ano-pVDZ & 25.35 & 17.65 & 23.99 & 16.64 \\
    &&&&&&&&&&aug-ano-pVDZ & 25.18 & 18.98 & 23.86 & 18.30 \\
    def2-TZVP & 16.17$^a$ & 10.76 & 15.41$^a$ & 9.68 & VTZ+d & 11.24 & 7.60 & 11.75  & 7.63 & ano-pVTZ & 10.66$^a$ & 7.17 & 10.09$^a$ & 6.68\\
    def2-TZVPP & 8.89 & 6.35 & 8.90 & 6.43 & haVTZ+d & 11.66 & 9.51 & 11.03 & 9.15 & saug-ano-pVTZ & 9.03 & 6.11 & 8.37 & 5.78 \\
    &&&&&&&&&&aug-ano-pVTZ & 8.83 & 6.58 & 8.62 & 6.40\\
    def2-QZVP & 4.13 & 2.64 & 5.75 & 3.94 & VQZ+d & 5.62 & 2.95 & 7.06 & 4.15 & ano-pVQZ & 5.10$^a$ & 3.34 & 5.89$^a$ & 4.13 \\
    def2-QZVPP & \multicolumn{4}{c|}{ditto} & haVQZ+d & 4.09 & 3.23 & 4.87 & 3.89 & saug-ano-pVQZ & 4.19 & 2.73 & 4.95 & 3.20 \\
    &&&&&&&&&&aug-ano-pVQZ & 3.62 & 2.48 & 4.51 & 2.87\\
                &    &      & & & V5Z+d   &      &      &      &      & ano-pV5Z      & 2.81 & 1.99 & 4.71 & 3.53 \\
                &    &      & & & haV5Z+d & 1.99 & 1.48 & 4.40 & 3.28 & saug-ano-pV5Z & 2.46 & 1.49 & 4.32 & 2.86 \\
    &&&&&&&&&&aug-ano-pV5Z & 2.52 & 1.26 & 4.51 & 2.08 \\
                \hline
                \multicolumn{4}{l}{CCSD(T*)(F12*)/}&&\multicolumn{4}{l}{CCSD(T)/}&&\multicolumn{4}{l}{CCSD(T*)(F12*)/}&\\
                cc-pVDZ-F12 & 2.75 & 2.00 & 5.42 & 4.01 & haV\{Q,5\}Z+d & 0.99 & 0.65 & 4.50 & 3.14 & cc-pVTZ-F12 & 0.70 & 0.50 & 4.69 & 3.63 \\

\hline
\tablenote{CBS refers to the CCSD(T) complete basis set limit}
\tablenote{$^a$ Using a step size of 0.005 a.u. during numerical differentiation leads to minor increases in RMSDs by 0.01-0.03 cm$^{-1}$.}
\tablenote{$^b$ 33.84 and 32.15 cm$^{-1}$, respectively, if the acetylene bending frequencies are excluded.}\\
\end{tabular}
\end{table}

First of all, concerning the reference, we could have made two basically equivalent choices, as  CCSD(T*)(F12*)/VQZ-F12 and pointwise CCSD(T)/haV\{Q,5\}Z+d extrapolation differ from each other by just 1.0 cm$^{-1}$ RMS. We have selected the former throughout. (For the avoidance of doubt, doubly and triply degenerate frequencies are assigned weights of 2 and 3, respectively, in the statistics.)

Second, the addition of tight $d$ functions to the second-row aug-cc-pVnZ basis sets has a very significant effect in SO$_2$ for smaller $n$. At first sight, no similar phenomenon is seen for the ANO basis sets; however, the primitive $\textit{d}$ functions that make up the $d$ symmetry ANOs have exponents 5.0755, 2.1833, 0.9392, 0.404, and 0.1738; hence, the high-exponent space is already adequately covered in the primitives.

Third, in both ANO and correlation consistent families, augmented basis sets have better statistics than unaugmented ones (with the exception of cc-pVDZ+d).

Fourth, among the ANO family, the fully augmented aug-ano-pVnZ basis sets have slightly better error statistics than the more economical `minimally augmented' saug-ano-pVnZ basis sets. 

Quadruple-$\zeta$ quality basis sets from all families outperform n = D and T members. The correlation-consistent haVQZ+d and the ANO set saug-ano-pVQZ yield similar RMSDs of 4.9 and 5.0 cm$^{-1}$ compared to experiment. For triple-$\zeta$ and especially double-$\zeta$, the ANO basis sets are markedly superior. Somewhat surprisingly, def2-TZVPP and def2-QZVP appear to be superior to their cc-pVnZ counterparts, and 
similar error reductions occur as more zetas are added, with n = Q having a small RMSD$_{exp}$ (5.8 cm$^{-1}$). (We note that def2-QZVP and def2-QZVPP are equivalent for the molecules considered here, and that the step-size choice for numerical differentiation, 0.01 or 0.005 a.u., has no significant effect on the calculated frequencies.)

CCSD(T) with def2-TZVP is in nearly thrice worse agreement with experiment than the complete basis set limit. Given that def2-TZVP is used in the high-level correction [CCSD(T)-MP2] because of its lower cost in the G4-type cWFTs, any improvements of the error statistics over `pure' CCSD(T)/def2-TZVP would make these cWFTs useful for spectroscopy.  

ANO basis sets outperform similarly-sized correlation consistent basis sets with notable \mbox{improvements} in RMSD$_{CBS}$, which are 7.2, 1.7, and 1 cm$^{-1}$ over VnZ with n = D, T, and Q, respectively.  The lowest errors are obtained for ano-pV5Z in CCSD(T) with an RMSD of 2.8 cm$^{-1}$ relative to reference. Another possibility is that the ANO basis sets – which are better equipped than small-medium sized correlation consistent basis sets for predicting accurately harmonic frequencies\cite{MTLbenzene,Schneider2008,McCaslin2013} – may offer advantages in composite wave function schemes as discussed below.

The hypersensitivity to the basis set of the acetylene bending frequencies was first noted by Lee and coworkers\cite{Simandiras1988} and analyzed in detail in Refs.\cite{jmlm104,Moran2006} as an intramolecular BSSE (basis set superposition error) problem. Another frequency that exhibits basis set hypersensitivity is the umbrella mode of ammonia --- which is a conspicuous outlier even for ano-pV5Z, less so for saug-ano-pV5Z.

\subsection{Composite wave function theory approaches}

Composite wave function theory has paved the way for cost-effective computations without significantly compromising accuracy. Various classes of additivity schemes have been studied previously, successfully predicting geometrical parameters,\cite{Heckert2005,Heckert2006} rotational constants,\cite{Puzzarini2008} and vibrational frequencies.\cite{Ruden2004,Morgan2018,Fortenberry2019,Gardner2021,Spiegel2023,Nelson2023}

The cost-effectiveness in these methods is achieved by combining various levels of electron correlation treatments using additive approximations, basis set extrapolations, and, when applicable, empirical corrections. By way of illustration, consider the following expression: 

E = MP2/LARGE + [CCSD(T)/SMALL – MP2/SMALL]

Now, if we express MP2 as HF + E2 and CCSD(T) as HF + E2 + HLC, we can simplify it further, as

E = HF/LARGE + E2/LARGE + HF/SMALL + E2/SMALL + HLC/SMALL – HF/SMALL – MP2/SMALL

where SMALL and LARGE correspond to two different basis set sizes. Simplifying this equation, we get 

E = HF/LARGE + E2/LARGE + HLC/SMALL. 

Therefore, the basis set for HF is effectively the same as LARGE for the MP2 correlation contribution, ensuring there is no mismatch as the other HF contributions cancel out. 

Table \ref{tab:rmsdshfreqcWFTs} details error statistics for harmonic frequencies in the HFREQ2014 species, comparing them to calculated and experimental data. Detailed equations of our G4-type composite schemes have been previously provided in Refs. \cite{Semidalas2020a,Semidalas2020} and the top performing G4-n, cc-G4-n, and G4-n-F12 methods in prediction of reaction energetics are now validated for harmonic frequencies. Some of these approaches are parameter-free while others achieve optimal results with a maximum of two fitted parameters. In addition, we consider the performance of W1$_{\rm val}$ and W2$_{\rm val}$ theories, i.e., Weizmann-n theories\cite{Martin1999,jmlm148,jmlm151} including only valence correlation. 

It can be rightly argued that, beyond small molecules where spectral inversion is comparatively easy, fundamental frequencies are more relevant for practical applications than harmonic frequencies. However, as shown in Table 4 of Ref.\cite{jmlm260}, even for CCSD(T)/CBS simple scaling of harmonic frequencies carries an intrinsic error of about 25 cm${-1}$, comparable to the uncertainty in hybrid DFT harmonic frequencies. The use of dual or multiple scaling factors for different frequency ranges at semi-arbitrary cutoff points (e.g.,\cite{Laury2011,Laury2012,ZapataTrujillo2022,ZapataTrujillo2023}) is a half-measure at best; second-order rotation-vibration perturbation theory (VPT2)\cite{Nielsen1951} will require a semidiagonal quartic force field. Schneider and Thiel\cite{Schneider1989} as far back as 1989 (in the context of semiempirical MO theory) pointed out that all the required force constants can be obtained by finite differences (in normal coordinates) of analytical second derivatives: this may be a viable approach for the present cWFT methods, or cWFT harmonic frequencies may be combined with DFT anharmonic force fields as demonstrated by Boese and Martin for the azabenzenes. 

\subsubsection{G4-type cWFTs based on CCSD(T)/def2-TZVP}

The most accurate result, using a def2 basis set for the CCSD(T) part, materializes for G4-T with RMSD$_{\rm CCSD(T)/CBS}$ of 4.7 cm$^{-1}$ and RMSD$_{\rm exp}$ of 5.4 cm$^{-1}$. In other words, not materially different from CCSD(T) at the valence CBS limit. Fundamentally, the same result is obtained if the force constants are obtained fully numerically at CCSD(T) and analytically at MP2. The lower-cost cost approach, G4-D, based on def2-SVP in CCSD(T), is three times worse in accuracy, while in G4-D-v2 with def2-SVPD, the error statistics are cut in half. The largest errors occur in the $\pi_g$ and $\pi_u$ degenerate bending modes of acetylene, which G4-D and G4-D-v2 underestimate by 26.9 and 39.9 cm$^{-1}$, while the accurate G4-T falls short by only 0.7 cm$^{-1}$. 

\begin{sidewaystable}
  \centering
  \caption{Root-mean-square deviations and mean absolute deviations (cm$^{-1}$) of calculated harmonic frequencies with various composite wave function schemes from computed CCSD(T*)(F12*)/VQZ-F12 calculations (referred to as CCSD(T)/CBS) and experiment for the HFREQ2014 dataset}
    \begin{tabular}{lccccrrrrrrcc}
    \hline
          &     \multicolumn{4}{c}{components}   &    \multicolumn{4}{c}{Errors relative to}\\
          
          &      & & &               &   \multicolumn{2}{c}{CCSD(T)/CBS}  & \multicolumn{2}{c}{Expt.} &   &          & \multicolumn{2}{l}{analytic derivatives\footnote{G: Gradient, H: Hessian, and E: Energy is doubly differentiated numerically.}}  \\\hline
    cWFT method & MP2 & N$_{bas}$ & CCSD(T) & N$_{bas}$ & RMSD & MAD & RMSD & MAD & c$_1$\footnote{c$_1$ is the extrapolation coefficient of MP2 correlation energies.} & 
    c$_2$=c$_3$\footnote{c$_2$ and c$_3$ are scaling parameters for E$_{CCSD-MP2}$ and E$_{(T)}$ terms, respectively.} & MP2   & CCSD(T) \\
 \hline
    G4-D-v0 & def2-\{SVSP,TZVPPD\} & \{36, 142\} & def2-SVSP & 36 & 26.07 & 21.99 & 27.29 & 22.62 & -0.07 & 1.05 & G & G \\
    G4-D    & def2-\{T,Q\}ZVPPD  & \{142, 258\}  & def2-SVP &  48 & 11.53  & 8.28 & 10.49 & 7.69 & 0.00  & 0.96  & G     & G \\
    G4-D-v2 & ditto & ditto   & def2-SVPD  &  72 &  8.20  & 6.13 & 7.83  & 5.83 & 0.00  & 0.96  & G     & G \\
    G4-T    & ditto & ditto   & def2-TZVP  & 86 &  4.65  & 3.32 & 5.41  & 4.35 & 0.57  & 1.05  & G     & G \\
    G4-T    & ditto & ditto   & def2-TZVP  & 86  &  4.62  & 3.34 & 5.37  & 4.23 & 0.57  & 1.05  & H     & E \\
\hline
    G4-D' & def2-\{T,Q\}ZVPPD' & \{130, 246\} & def2-SVP & 48 & 11.54 & 8.32 & 10.52 & 7.71 & 0.00  & 0.96  & G     & G \\
    G4-D'-v2 & ditto & ditto   & def2-SVPD  & 72 &   8.23 & 6.16 & 7.87  & 5.86 & 0.00  & 0.96  & G     & G \\
    G4-T'    & ditto & ditto   & def2-TZVP  & 86 &   4.74 & 3.45 & 5.49  & 4.42 & 0.57 & 1.05  & G     & G \\
    G4-T'    & ditto & ditto   & def2-TZVP  & 86 &   4.73 & 3.46 & 5.53  & 4.40 & 0.57  & 1.05  & H     & E \\
\hline
    cc-G4(FC)-D    & awCV\{T,Q\}Z       &  \{210,402\}    & def2-SVP & 48 & 13.54 & 9.93 & 12.95 & 9.81  & 1.09  & 1.12  & G     & G \\
    cc-G4-D        & awCV\{T,Q\}Z + CV  &  ditto   & def2-SVP  & 48 & 11.49 & 7.80 & 11.29 & 7.84  & 1.09  & 1.12  & G     & G \\
    cc-G4(FC)-D-v2 & awCV\{T,Q\}Z       &  ditto   & def2-SVPD & 72 & 9.10 & 7.06 & 9.28  & 7.07  & 1.09  & 1.12  & G     & G \\
    cc-G4-D-v2 & awCV\{T,Q\}Z + CV      &  ditto   & def2-SVPD & 72 & 6.91 & 4.70 & 7.77  & 5.56  & 1.09  & 1.12  & G     & G \\
    cc-G4(FC)-T    & awCV\{T,Q\}Z       &  ditto   & def2-TZVP & 86 &  5.30  & 4.24  & 5.45 & 4.36  & 0.63  & 1.03  & G     & G \\
    cc-G4-T    & awCV\{T,Q\}Z + CV      &  ditto   & def2-TZVP & 86 &  4.25  & 3.05  & 5.07 & 3.84   & 0.63  & 1.03  & G     & G \\
\hline
    G4-T$_{\rm ano}$-v1 & ano-pVQZ & 230 & ano-pVTZ  & 116 &  6.65  & 4.49 & 6.55 & 4.65 & -     & -     & G     & G \\
    G4-T$_{\rm ano}$-v2 & ano-pV5Z & 302 & ano-pVTZ  & 116 &  4.98  & 3.75 & 5.14 & 3.93  & -     & -     & G     & G \\
\hline
    cWFT method & MP2-F12 & N$_{bas}$ & CCSD(T) & N$_{bas}$ &   RMSD & MAD & RMSD & MAD & c$_1$ & c$_2$=c$_3$ &  MP2-F12 & CCSD(T)  \\
\hline
    G4-T$_{\rm ano}$-F12-v0 & VDZ-F12 & 96 & ano-pVTZ  & 116 & 10.92  & 7.73 & 9.94   & 7.84  &    -   & -     & G     & G \\
    G4-D$_{\rm ano}$-F12-v0 & VDZ-F12 & 96 & ano-pVDZ  & 48 &  13.54  & 9.94 & 12.27  & 9.58  &    -   & -     & G     & G \\
    G4-T$_{\rm ano}$-F12-v1 & VTZ-F12 & 178 & ano-pVTZ  & 116 &  4.97   & 3.56 & 5.21   & 3.93   &    -   & -     & G     & G \\
    G4-D$_{\rm ano}$-F12-v1 & VTZ-F12 & 178 & ano-pVDZ  & 48 &  7.99   & 5.72 & 7.36   & 5.62   &   -    & -     & G     & G \\
    G4-T$_{\rm ano}$-F12-v2 & VQZ-F12 & 310 & ano-pVTZ  & 116 &  3.74   & 2.93 & 4.58   & 3.45  &   -    & -     & G     & G \\
    G4-D$_{\rm ano}$-F12-v2 & VQZ-F12 & 310 & ano-pVDZ  &  48 &  7.04   & 5.29 & 6.72   & 5.16   &   -    & -     & G     & G \\
\hline
    W1 (Val.)               &           &       &  &  & 1.96    & 1.40 & 4.76   & 3.54   &        & & G     & G \\
        W2 (Val.)           &           &       &  &  & 1.15    & 0.70 & 4.67   & 3.26   &        & & G     & G \\
\hline  
    \end{tabular}%
  \label{tab:rmsdshfreqcWFTs}%

\end{sidewaystable}

An important technical note for wave function calculations using basis sets augmented with diffuse functions, particularly for linear molecules, is that employing such basis sets, e.g. def2-nZVPPD (n=T, Q) for acetylene, may result in an overlap matrix S with very small eigenvalues. Gaussian implements a form of SVD (singular value decomposition) in which the eigenvectors of S with eigenvalues below a cutoff (default value: 10$^{-6}$) are discarded. This leads to truncation of the virtual orbital space, which may not be consistent across the surface — or even along a single normal mode displacement — and hence may cause erratic harmonic frequencies. In the present work, this occurred for the bending frequencies of acetylenes.  To address this, we disabled the 'SVD screening'  in Gaussian using \texttt{IOp(3/32)=2}; MOLPRO has no such screening in the first place, but for acetylene, def2-nZVPPD led to S eigenvalues below the default \texttt{THROVL}=$10^{-8}$, and lowering THROVL brought on numerical issues that required severely tightening the integral evaluation cutoffs. No such problems were seen if the 'D' functions were retained on the heavy atoms but not on H, which we denote def2-nZVPPD$'$ and applied for acetylene.

Substituting $E_{\rm 2/def2-\{T,Q\}ZVPD'}$ in G4 type approaches, where diffuse functions are omitted on hydrogen atoms, has no significant effect on error statistics when compared to reference or experiment. G4-T' stands out as the most accurate with an RMSD$_{\rm CBS}$ of  4.7$_4$ cm$^{-1}$, a mere 0.11 cm$^{-1}$ above regular G4-T. Moreover, opting for def2-{T,Q}ZVPD' in G4-type approaches effectively addresses concerns related to small eigenvalues in the overlap matrix, particularly for linear molecules, without compromising accuracy.

In our initial studies on G4-like cWFTs\cite{Semidalas2020a,Semidalas2020} we did not explore a \{D,T\} extrapolation in MP2, like E2/def2-\{SVSP,TZVPPD\} +[CCSD(T)-MP2]/def2-SVSP, where def2-SVSP means no p functions on hydrogen atoms, assuming it would be less accurate than other cWFTs. We trained such a G4-D-v0 composite on the GMTKN55 dataset and found a high WTMAD2 (weighted mean absolute deviation) value of 4.77 kcal/mol compared to CCSD(T)/CBS data, extracted from the ACCDB database,\cite{Morgante2019} or higher level from our earlier work. The RMSD values for harmonic frequencies are unacceptably high, reaching 26.07 and 27.29 cm$^{-1}$ compared to CCSD(T)/CBS and experimental values: this is on par with (much cheaper) hybrid DFT functionals such as B3LYP and TPSS0.\cite{jmlm260} However, standard CCSD(T)/def2-SVSP is far less accurate than G4-D-v0, with corresponding RMSDs of 61.37 and 60.46 cm$^{-1}$. Clearly, there is no advantage in a \{D,T\} extrapolation and if 10 cm$^{-1}$ errors are acceptable then empirical spin-scaled double-hybrid functionals, such as DSD-PBEP86-D3(BJ)\cite{Kozuch2013} and revDSD-PBEP86-D4\cite{Santra2019a}, represent a much more cost-effective alternate.

Additionally, we examined the impact of using VDZ-F12 in MP2-F12 instead of VTZ-F12. We saw RMSD values relative to CCSD(T)/CBS rise high as 13.54 and 10.92 cm$^{-1}$ with MP2-F12/VDZ-F12 and either [CCSD(T)-MP2]/ano-pVDZ or ano-pVTZ, respectively, which is on par with the performance of double hybrid functionals.

What about the effects of core-core and core-valence correlation? To answer that, we consider the correlation consistent augmented aug-cc-pwCVnZ basis sets that provide the necessary radial and angular flexibility in the core-valence region. In 2018, Sylvetsky and Martin\cite{jmlm281} showed that a awCV\{T,Q\}Z basis set extrapolation at CCSD(T) proved sufficient and captured the most significant part of electron correlation. Here, we follow a two-step approach to assess those effects. Firstly, we check the effects of increased basis set radial flexibility for {\em valence} correlation by replacing def2-\{T,Q\}ZVPPD with awCV\{T,Q\}Z at the MP2 level, while only correlating the valence electrons. In doing so, no material gains in accuracy are seen, but rather the reverse. The RMSD$_{\rm CCSD(T)/CBS}$ increases by 2 cm$^{-1}$ for cc-G4(FC)-D, 0.9 cm$^{-1}$ for cc-G4(FC)-D-v2, and 0.6 cm$^{-1}$ for cc-G4-T, each compared to the corresponding G4-n. Secondly, we correlate all electrons in MP2 and the results showed the CV correlation improves the RMSD by 2.0 cm$^{-1}$ in cc-G4-D relative to cc-G4(FC)-D, with the former achieving identical accuracy to G4-D. Further improvement of 2.2 cm$^{-1}$ occurs with def2-SVPD in CCSD(T). The lowest RMSD of 4.25 cm$^{-1}$ is found for cc-G4-T, this result represents a 1.0 cm$^{-1}$ amelioration over valence-only cc-G4(FC)-T, and even surpassing G4-T by 0.3 cm$^{-1}$.

For higher accuracy regimes, we refer the reader to our recent study\cite{Spiegel2023} on ground-state spectroscopic constants of diatomic molecules from post-CCSD(T) up to CCSDTQ56. We showed there that 2 cm$^{-1}$ accuracy is achievable on a semi-routine basis (see Table 5 in ref. \cite{Spiegel2023}), but that this requires {\em both} post-CCSD(T) valence correlation correction at least at the CCSDT(Q)$_\Lambda$ level {\em and} core-valence correlation corrections at the CCSD(T) level. (Including each on its own will actually make agreement {\em worse}, as valence CCSD(T) benefits from a felicitous error compensation.)
We have repeated the Dunham analyses\cite{Dunham1932} from Ref.\cite{Spiegel2023} without the scalar relativistic correction. The largest individual difference in $\omega_e$ is seen for HCl (-4.3 cm$^{-1}$) followed by -2.8 cm$^{-1}$ for HF, but most effects are on the order of 1 cm$^{-1}$ or less. Thus, the RMSD on $\omega_e$ increased only mildly, from 2.10 to 2.55 cm$^{-1}$, while the effect on other spectroscopic constants was negligible: obviously, compared to 5-10 cm$^{-1}$ RMSDs for more approximate methods, such an increase is entirely negligible. Needless to say, this will no longer be the case for heavy p-block compounds. 

We attempted to add diagonal Born-Oppenheimer corrections at the CCSD/aug-cc-pVTZ level using the implementation\cite{DBOC_CCSD} in CFOUR\cite{CFOUR}. While corrections may exceed 1 cm$^{-1}$ for H$_2$, BH, and the like, for heavier diatomics they are negligible compared to other remaining error sources.

\begin{figure}[h]
\label{fig:hfreq_cWFTs}
\caption{Box-and-whisker plot for harmonic frequency deviations of composite wave function schemes from CCSD(T*)(F12*)/VQZ-F12 for the HFREQ2014 dataset. Plot description details are the same as for Fig. \ref{fig:hfreq_CCSDt}.}
\includegraphics[width=5in]{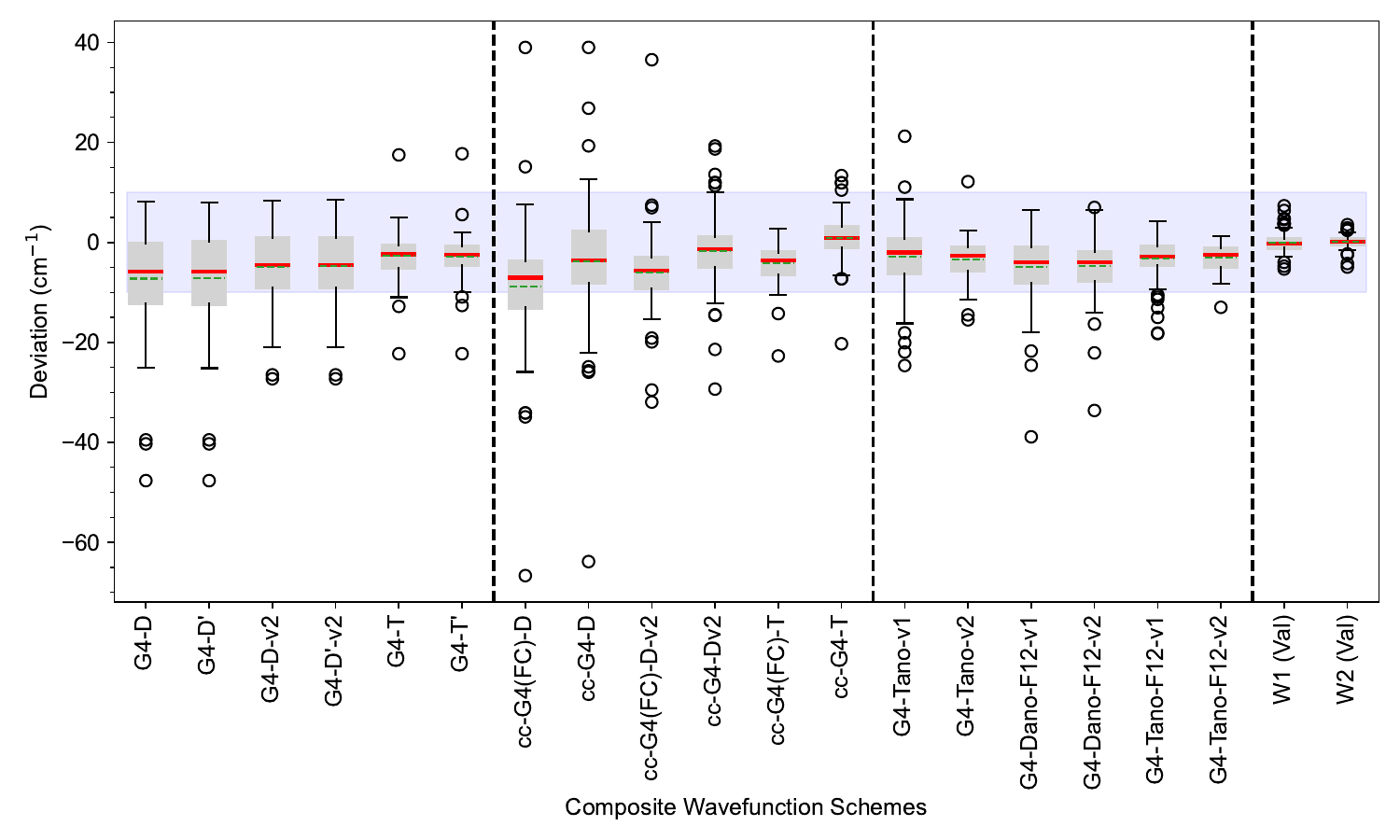}
\end{figure}

\subsubsection{ANO basis sets for G4-type cWFTs}

ano-pVnZ basis sets are now considered in the CCSD(T) part of cWFTs for the harmonic frequencies in HFREQ2014. The energy expressions of these cWFTs are simply derived from the sum of the total MP2 energy with a larger basis set and the higher level [CCSD(T)-MP2]/ano-pVTZ correction, without introducing empirical parameters. Using E$_{2/ano-pVQZ}$ in G4-T$_{\rm ano}$-v1, we find an RMSD of 6.6 cm$^{-1}$ w.r.t both reference and experiment (see Table \ref{tab:rmsdshfreqcWFTs} and Fig. \ref{fig:hfreq_cWFTs}). Closer agreement to CCSD(T) at the complete basis set limit is achieved with ano-pV5Z in MP2, leading to the G4-T$_{\rm ano}$-v2 variant, which shows better RMSD$_{exp}$ than G4-T' and G4-T by by approximately 0.3 cm$^{-1}$.
 
In our earlier work,\cite{Semidalas2020} we found that MP2-F12-based methods, such as the parameter-free cc-G4-F12-T, yielded the lowest WTMAD2 values ($\sim$1.0 kcal/mol) for the energetics of the GMTKN55 benchmark suite.\cite{Goerigk2017a} That prompts the question 
whether the predicted harmonic frequencies can become more accurate  by substituting explicitly correlated \mbox{MP2-F12} for conventional MP2 in composite energy schemes. Among the tested MP2-F12-based variants, \mbox{G4-T$_{\rm ano}$-F12-v2} is the most accurate with an RMSD of 3.7 cm$^{-1}$ relative to reference calculated frequencies, compared to 7 cm$^{-1}$ for G4-D$_{\rm ano}$-F12-v2. Reducing the basis set size to n = T from n = Q in MP2-F12/VnZ-F12 worsens RMSD$_{\rm CCSD(T)/CBS}$ by 1.2 cm$^{-1}$ for \mbox{G4-T$_{\rm ano}$-F12} and by 0.9$_5$ cm$^{-1}$ for G4-D$_{\rm ano}$-F12.  Clearly, combining MP2-F12 correlation with CCSD(T)/ano-pVnZ presents an attractive option for accurate vibrational frequencies in parameter-free cWFTs. The calculated frequencies can be inspected in the Supporting Information.

Finally, we note that the Weizmann-n methods, W1 and W2, lead to the lowest RMSDs relative to the calculated reference harmonic frequencies, at 1.96 and 1.15 cm$^{-1}$, respectively. There is no systematic improvement in RMSD$_{exp}$ values over the most accurate ANO-based method, indicating that the convergence towards CCSD(T)/CBS has been achieved.

\subsection{Harmonic frequencies of benzene: Successes and limitations of cWFTs}
In 1997, Martin, Taylor, and Lee\cite{MTLbenzene} computed the CCSD(T) geometry and  harmonic force field of C$_6$H$_6$, and noted that the two out-of-plane ring modes $\omega_4$ and $\omega_5$ exhibit a more pronounced form of the same hypersensitivity as seen for the acetylene bending frequencies\cite{Simandiras1988} and traced to intramolecular BSSE in Ref.\cite{jmlm104}. (Moran et al.\cite{Moran2006} later extended this discussion to benzene.) Limitations of available computers at the time precluded going to larger basis sets such as haVQZ, but since ANOs minimize the BSSE for a given contracted size, ANO4321 was attempted and found to be resilient than cc-pVTZ.

As extracting a full set of experimental harmonic frequencies and anharmonicity constants for such a large molecule would require a staggering number of vibrational band origins, the available `experimental' harmonic frequencies of Miani et al.\cite{Miani2000} are in truth semi-experimental (a term introduced in the spectroscopic realm by Jean Demaison\cite{Demaison2007}), namely, from combining experimental fundamentals with a DFT calculated quartic force field. (We note in passing that results of a `blind challenge' on the ground-state correlation energy of benzene were recently reported.\cite{Eriksen2020}.)

Hence benzene would appear to be a good `proof of concept' for the application of composite WFTs to harmonic frequencies of not-so-small molecules. Owing to the high symmetry, we can actually carry out CCSD(T) and CCSD(T*)(F12*) calculations close to the basis set limit, giving us a realistic reference. Any cWFT that would reproduce the harmonic frequencies well might be a good candidate for an anharmonic force field.

Table \ref{tab:rmsdsbenzene} showcases the calculated harmonic frequencies using ‘pure’ coupled-cluster methods and their corresponding cWFTs. For the reference, we consider CCSD(T)/ano-pV5Z harmonic frequencies. Geometry optimizations were carried out in D$_{2h}$ point group symmetry and the Hessian was obtained through the method of finite differences.

Overall, the RMSDs consistently improve with larger basis sets, ranging from 3.6 cm$^{-1}$ for CCSD(T)/ano-pVTZ to 1.5 cm$^{-1}$ for ano-pVQZ. CCSD(T*)(F12*)/cc-pVTZ-F12 has an RMSD of 4.5 cm$^{-1}$, which drops to 1.6 cm$^{-1}$ for cc-pVQZ-F12. Much of that is due to the two problematic $\omega_4$ and $\omega_5$ modes, however: conspicuous discrepancies of -17 and -14 cm$^{-1}$, respectively, are observed at the CCSD(T*)(F12*)/VTZ-F12 level; these however are drastically reduced to  \{-7,-3\} cm$^{-1}$,  for VQZ-F12, and still further to \{-3,+1\} cm$^{-1}$ for V\{T,Q\}Z-F12 extrapolation.
Statistics without them are RMSD = 1.9 for VTZ-F12 and 0.9 cm$^{-1}$ for VQZ-F12, which is more in line with HFREQ2014.
 A \{T,Q\} extrapolation (with an exponent of  4.5960, Table X in Hill et al.\cite{Hill2009a}) yields RMSD=1.0 cm$^{-1}$ including all modes, and 0.8 cm$^{-1}$ excluding $\omega_4$ and $\omega_5$.  

Next, we estimate the accuracy of composite schemes for benzene. The most accurate results are obtained in parameter-free methods based on a high level correction [CCSD(T)-MP2/ano-pVTZ] that is combined with either E$_{\rm 2/ano-pVQZ}$ (RMSD = 3.0 cm$^{-1}$) or E$_{\rm 2/ano-pV5Z}$ (RMSD = 2.8 cm$^{-1}$). Similar gains are observed for the MP2-F12-based G4-T$_{\rm ano}$-F12-v2, which exhibits an RMSD of 2.6 cm$^{-1}$, and this result is 1.0 cm$^{-1}$ lower than plain  CCSD(T)/ano-pVTZ. Scaling the triples term $E_{\rm (T)}$ does more harm than good, resulting in an increase of $\sim$0.2 cm$^{-1}$ for combinations of MP2-F12/VnZ-F12 with CCSD(T)/ano-pVTZ. 

For the low-cost G4-T, with an RMSD of 6.4 cm$^{-1}$, the ${\omega}_4$ and ${\omega}_5$ modes are underestimated by -27 and -8 cm$^{-1}$, respectively. Introducing MP2-F12 correlation, instead of conventional MP2, leads to a slight deterioration of 0.2 cm$^{-1}$ with VTZ-F12, but a significant improvement of 2.0 cm$^{-1}$ occurs with VQZ-F12.

Consequently, for accurate harmonic frequencies it is recommended to combine MP2-F12/VnZ-F12 (n=T or Q) with CCSD(T)/ano-pVTZ, as such parameter-free cWFTs offer substantial gains.

\begin{table}[]
\caption{Calculated and experimentally derived harmonic frequencies  (in cm$^{-1}$) for the benzene molecule}
\hspace{-1.5cm}
\small
\begin{tabular}{llrrrr|rrr|ccc}
\hline
\hline
&&Expt.&\multicolumn{3}{c|}{CCSD(T)/}& \multicolumn{3}{c|}{CCSD(T*)(F12*)/} & \multicolumn{3}{c}{MP2-F12/}\\
&&Ref.&ano-pV5Z&ano-pVQZ&\multicolumn{1}{c|}{ano-pVTZ}&V\{T,Q\}Z-F12&VQZ-F12&VTZ-F12&VDZ-F12&VTZ-F12&VQZ-F12\\
&&\cite{Miani2000}&&    &        &       &       &    &\multicolumn{3}{c}{+[CCSD(T)-MP2]/def2-TZVP}\\
\hline
$\omega_1$    & a$_{1g}$   & 1010.0      & 1006.6   & 1005.5   & 1003.3 & 1006.3 & 1006.5 & 1007.1 & 1006.3  & 1007.8 & 1005.6                                             \\
$\omega_2$    & a$_{1g}$   & 3218.0      & 3204.3   & 3208.0   & 3209.9 & 3205.0   & 3205.4 & 3206.5 & 3209.0  & 3208.9 & 3204.9                                              \\
$\omega_3$    & a$_{2g}$   & 1392.0      & 1380.5   & 1378.3   & 1374.9 & 1382.8 & 1382.0   & 1379.7 & 1374.0  & 1378.1 & 1380.1                                              \\
$\omega_4$    & b$_{2g}$   & 717.0       & 712.4    & 711.6    & 708.4  & 709.1 & 705.4  & 695.2   & 679.7   & 687.4  & 695.7                                                \\
$\omega_5$    & b$_{2g}$   & 1012.0      & 1012.4   & 1010.1   & 1006.8 & 1013.4 & 1009.5 & 998.7  & 977.5   & 992.4  & 998.9                                                \\
$\omega_6$    & e$_{2g}$   & 617.0       & 611.5    & 611.0    & 610.3  & 611.8 & 611.8 & 611.7    & 610.4   & 610.6  & 610.6                                                 \\
$\omega_7$    & e$_{2g}$   & 3210.0      & 3179.6   & 3181.5   & 3184.6 & 3180.6 & 3180.9 & 3181.8 & 3184.1  & 3183.9 & 3179.8                                               \\
$\omega_8$    & e$_{2g}$   & 1645.0      & 1639.3   & 1638.5   & 1637.0 & 1639.5 & 1639.6 & 1639.9 & 1640.2  & 1640.6 & 1638.4                                              \\
$\omega_9$    & e$_{2g}$   & 1197.0      & 1191.8   & 1191.4   & 1192.3 & 1192.1 & 1192.1 & 1192.2 & 1192.5  & 1190.5 & 1191.1                                                \\
$\omega_{10}$ & e$_{1g}$   & 861.0       & 863.4    & 863.1    & 863.6  & 863.5 & 862.7 & 860.5    & 856.4   & 857.6  & 860.4                                                  \\
$\omega_{11}$ & a$_{2u}$   & 683.0       & 683.6    & 683.9    & 685.9  & 683.6 & 683.2 & 682.1    & 676.3   & 678.6  & 681.1                                                   \\
$\omega_{12}$ & b$_{1u}$   & 1030.0      & 1024.6   & 1022.9   & 1019.4 & 1025.7 & 1025.4 & 1024.5 & 1022.3  & 1022.1 & 1022.4                                                \\
$\omega_{13}$ & b$_{1u}$   & --          & 3169.9   & 3171.4   & 3174.4 & 3171.1 & 3171.4 & 3172.3 & 3173.2  & 3174.4 & 3170.3                                                \\
$\omega_{14}$ & b$_{2u}$   & 1338.0      & 1328.6   & 1329.1   & 1328.3 & 1329.1 & 1329.7 & 1331.4 & 1319.9  & 1323.9 & 1321.1                                                 \\
$\omega_{15}$ & b$_{2u}$   & 1163.0      & 1158.4   & 1158.5   & 1160.5 & 1158.4 & 1158.5 & 1158.9 & 1158.0  & 1156.3 & 1156.9                                                 \\
$\omega_{16}$ & e$_{2u}$   & 406.0       & 406.3    & 406.4    & 405.8  & 405.6 & 405.3 & 404.6 & 399.6   & 403.2  & 403.5                                                    \\
$\omega_{17}$ & e$_{2u}$   & 987.0       & 987.0    & 985.3    & 983.0  & 987.2 & 985.9 & 982.3  & 977.5   & 984.0  & 986.0                                                  \\
$\omega_{18}$ & e$_{1u}$   & 1057.0      & 1057.2   & 1056.2   & 1054.8 & 1057.4 & 1057.4 & 1057.5  & 1055.0  & 1057.0 & 1056.2                                                 \\
$\omega_{19}$ & e$_{1u}$   & 1522.0      & 1511.3   & 1509.5   & 1506.6 & 1512.6 & 1512.5 & 1512.2 & 1513.2  & 1511.6 & 1511.2                                                 \\
$\omega_{20}$ & e$_{1u}$   & 3212.0      & 3195.1   & 3197.9   & 3200.7 & 3195.9 & 3196.3 & 3197.3 & 3199.2  & 3199.8 & 3195.5                                                 \\
\hline
     & \multicolumn{2}{c}{RMSD (cm$^{-1}$)} & REF   & 1.53     & 3.62   & 0.99 & 1.64   & 4.45 & 9.93   & 6.65  & 4.38\\
     & \multicolumn{2}{c}{without $\omega_4,\omega_5$} & REF   & 1.52     & 3.51   & 0.79 & 0.90   & 1.99 & 4.90 &  3.26  & 2.02\\     \hline
\end{tabular}
  \label{tab:rmsdsbenzene}%
\end{table}

\begin{table}[]
\addtocounter{table}{-1}
\caption{(continued)}
 \hspace{-1.5cm}
\small
\begin{tabular}{rrr|rrr|rr}
\hline
    &&&MP2-F12/&MP2-F12/&MP2-F12/&MP2/&MP2/\\
    &&G4-T&VDZ-F12&VTZ-F12&VQZ-F12&ano-pVQZ&ano-pV5Z\\
    &&&\multicolumn{3}{c|}{+[CCSD(T)-MP2]/ano-pVTZ$^a$}&\multicolumn{2}{c}{+[CCSD(T)-MP2]/ano-pVTZ}\\
    \hline
$\omega_1$  & a$_{1g}$         & 1018.0 & 1006.3 (0.1)                                                          & 1007.8 (0.1)                                                          & 1009 (-0.6)                                                           & 1007.8                                                                & 1010.1                                                                \\
$\omega_2$  & a$_{1g}$         & 3201.4 & 3209.8 (-0.6)                                                         & 3209.7 (-0.6)                                                         & 3205.7 (-0.6)                                                         & 3207.7                                                                & 3202.7                                                                \\
$\omega_3$  & a$_{2g}$         & 1380.0 & 1373.3 (0)                                                            & 1377.4 (0)                                                            & 1379.4 (0)                                                            & 1376.3                                                                & 1378.4                                                                \\
$\omega_4$  & b$_{2g}$         & 738.8 & 685.4 (-0.6)                                                          & 695.5 (-0.6)                                                          & 703.4 (-0.7)                                                          & 709.5                                                                 & 710.1                                                                 \\
$\omega_5$  & b$_{2g}$         & 1004.5 & 986.5 (0)                                                             & 1002.3 (-0.6)                                                         & 1005.6 (2.9)                                                          & 1004.4                                                                & 1005.0                                                                \\
$\omega_6$  & e$_{2g}$         & 608.7 & 610.7 (0)                                                             & 610.9 (0)                                                             & 610.9 (0)                                                             & 610.1                                                                 & 610.5                                                                 \\
$\omega_7$  & e$_{2g}$         & 3172.9 & 3185 (-0.7)                                                           & 3184.8 (-0.7)                                                         & 3180.6 (-0.8)                                                         & 3180.7                                                                & 3177.5                                                                \\
$\omega_8$  & e$_{2g}$         & 1636.7 & 1639.8 (0.4)                                                          & 1640.2 (0.4)                                                          & 1638 (0.4)                                                            & 1637.1                                                                & 1637.1                                                                \\
$\omega_9$  & e$_{2g}$         & 1190.3 & 1192.2 (-0.2)                                                         & 1190.2 (-0.2)                                                         & 1190.8 (-0.2)                                                         & 1190.3                                                                & 1190.4                                                                \\
$\omega_{10}$ & e$_{1g}$       & 861.3 & 857.8 (-0.6)                                                          & 859 (-0.6)                                                            & 861.8 (-0.6)                                                          & 861.7                                                                 & 861.9                                                                 \\
$\omega_{11}$ & a$_{2u}$       & 680.9 & 681 (-0.6)                                                            & 680 (-0.6)                                                            & 682.4 (-0.6)                                                          & 682.9                                                                 & 682.5                                                                 \\
$\omega_{12}$ & b$_{1u}$       & 1028.8 & 1023 (-0.3)                                                           & 1022.9 (-0.3)                                                         & 1023.1 (-0.3)                                                         & 1021.0                                                                & 1022.5                                                                \\
$\omega_{13}$ & b$_{1u}$       & 3167.0 & 3174.1 (-0.7)                                                         & 3175.2 (-0.7)                                                         & 3171.2 (-0.7)                                                         & 3170.5                                                                & 3167.8                                                                \\
$\omega_{14}$ & b$_{2u}$       & 1322.1 & 1323.7 (-4.8)                                                         & 1327.7 (-4.9)                                                         & 1324.9 (-4.8)                                                         & 1321.7                                                                & 1319.6                                                                \\
$\omega_{15}$ & b$_{2u}$       & 1155.4 & 1157.7 (-0.9)                                                         & 1156 (-0.9)                                                           & 1156.6 (-0.9)                                                         & 1156.6                                                                & 1156.0                                                                \\
$\omega_{16}$ & e$_{2u}$       & 401.7 & 400 (-0.5)                                                            & 403.5 (-0.5)                                                          & 403.8 (-0.5)                                                          & 405.3                                                                 & 404.9                                                                 \\
$\omega_{17}$ & e$_{2u}$       & 989.5 & 976.6 (-0.5)                                                          & 983.2 (-0.4)                                                          & 985.2 (-0.4)                                                          & 983.4                                                                 & 985.0                                                                 \\
$\omega_{18}$ & e$_{1u}$       & 1055.5 & 1055 (-0.1)                                                           & 1057 (-0.1)                                                           & 1056.2 (-0.1)                                                         & 1055.1                                                                & 1055.7                                                                \\
$\omega_{19}$ & e$_{1u}$       & 1508.9 & 1512.3 (0.2)                                                          & 1510.7 (0.2)                                                          & 1510.3 (0.2)                                                          & 1507.6                                                                & 1509.1                                                                \\
$\omega_{20}$ & e$_{1u}$       & 3191.2 & 3200 (-0.7)                                                           & 3200.6 (-0.7)                                                         & 3196.3 (-0.7)                                                         & 3197.3                                                                & 3193.2                                                                \\
 \hline                                       
\multicolumn{2}{c}{RMSD (cm$^{-1}$)} &  6.37 & 8.18 (0.25) & 4.77 (0.22) & 2.57 (0.31) & 2.98                                                                   & 2.82    \\      
\multicolumn{2}{c}{without $\omega_4,\omega_5$} &  4.04 & 4.67 (0.32) & 3.25 (0.16) & 1.58 (0.65) & 2.63                                                                   & 2.52    \\      \hline    
\end{tabular}
\tablenote{(a) Values in parentheses indicate deviations between cWFTs having their [CCSD(T)-MP2] term scaled by 1.04382 and parameter-free cWFTs (c$_{[CCSD(T)-MP2]}$ = 1).}
\end{table}

\subsection{Timing comparison of cWFT and DFT methods}
\begin{figure}[h!]
\label{fig:hfreq_timings}
\caption{Visual representation of wall clock times for selected cWFT and DFT methods. Note that the y axis is logarithmic.}
\includegraphics[width=4in]{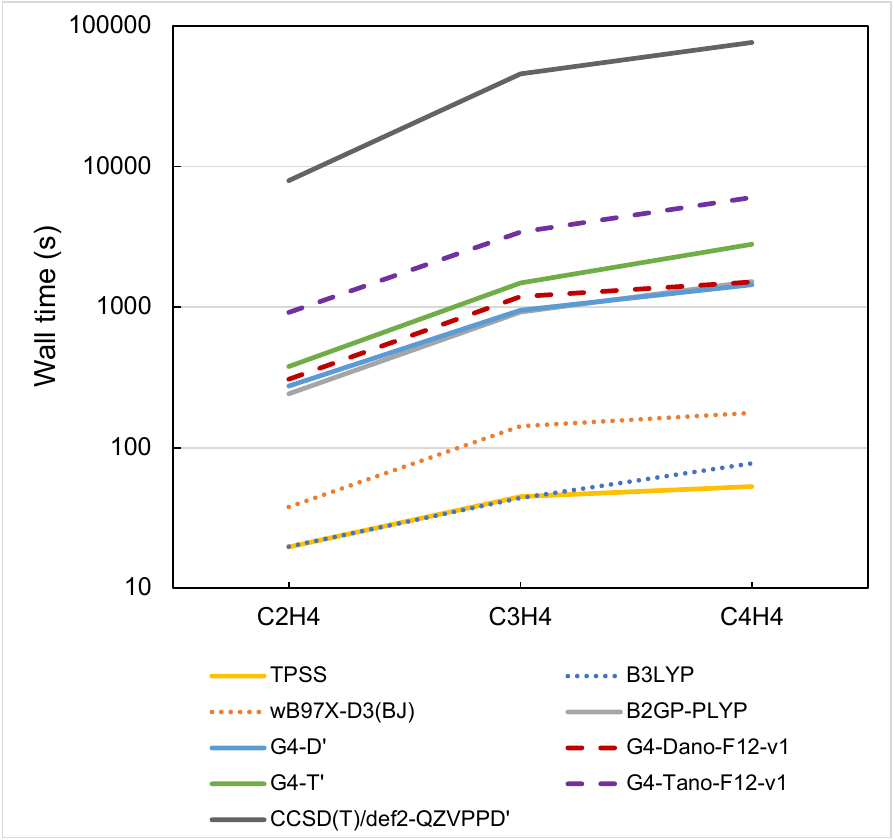}
\end{figure}

The reviewers requested a timing comparison of the present and alternative approaches. Fig. \ref{fig:hfreq_timings} depicts ‘wall clock’ times of selected cWFT and DFT methods for semi-numerical evaluation of harmonic frequencies obtained using MOLPRO from non-stationary geometries. All calculations ran on identical architecture nodes using 16 physical cores of an Intel Xeon Gold 5320 CPU with a maximum memory of 46.4 GB per core.

A notable speedup occurs with increasing molecular size when using cWFTs compared to 'brute force' CCSD(T) in the largest basis set of the composite scheme, the latter achieves speedups for cyclobutadiene by a factor of 53 for G4-D' and 27 for G4-T'. For  ethylene, those ratios decrease to 29 and 21, respectively.

The MP2-F12-based method G4-D$_{ano}$-F12-v1 is on par with G4-D’. However, the former exploits the accelerated basis set convergence at the MP2-F12 level, resulting to a  3.55 cm$^{-1}$ improvement in RMSD for HFREQ2014. Moreover, using the triple-$\zeta$ basis set VTZ-F12 in MP2-F12 makes G4-T$_{ano}$-F12-v1 nearly three times more expensive than with VDZ-F12.  Despite the increased cost, G4-T$_{ano}$-F12-v1 is almost an order of magnitude less expensive than CCSD(T)/def2-QZVPPD'.

To conclude, we assess the computational costs of DFT for harmonic frequencies alongside cWFTs; in order to keep the timing comparison fair, we carried out these calculations using MOLPRO. Specifically, as representative examples for timing, we picked the meta-GGA functional TPSS\cite{Tao2003} and the hybrid functionals B3LYP\cite{Lee1988a,Becke1993a} and $\omega$B97X-D3(BJ),\cite{Lin2013} as well as the double-hybrid B2GP-PLYP,\cite{Karton2008} which will have a similar cost as DSD-PBEP86-D3(BJ)\cite{Kozuch2013} or revDSD-PBEP86-D4\cite{Santra2019a} (which presently do not have analytical gradients in MOLPRO). Notably, G4-D’ and B2GP-PLYP exhibit similar costs. Hybrid GGA functionals, such as $\omega$B97X-D3(BJ) and $\omega$B97X-V\cite{Mardirossian2014} present the lowest-cost option, if ca. 30 cm$^{-1}$ accuracy is acceptable.

\section{Conclusions}

In this study, we have thoroughly examined the performance of various coupled-cluster composite wave function approaches (cWFT) for harmonic frequencies. Our investigation involved the development of extrapolation formulas for force constants, while also enabling geometry optimization and harmonic frequency calculations through an implementation we provide.

We have validated the top-performing composite energy schemes, based on previous evaluations for reaction energetics in Refs. \cite{Semidalas2020,Semidalas2020a} using the large GMTKN55 test suite, against the harmonic frequencies in the HFREQ2014 dataset from CCSD(T*)(F12*)/VQZ-F12 calculations and experimental data. G4-T is three times more accurate than plain CCSD(T)/def2-TZVP, while G4-T$_{\rm ano}$ is twice as accurate as CCSD(T)/ano-pVTZ. Notably, ANO basis sets combined with explicitly correlated MP2-F12, such as G4-T$_{\rm ano}$-F12, show promising performance, achieving accuracy of ~5 cm$^{-1}$ compared to the experiment, and they are on par with the accuracy of CCSD(T) at the complete basis set limit. Following closely was our standard G4-T approach, built upon def2-TZVP for the high level correction. Additionally, the Weizmann-n theories, W1 and W2, delivered the most accurate results when compared to the calculated reference harmonic vibrational frequencies.

The addition of diffuse functions on hydrogen does not materially help performance for neutral molecules, and in fact causes significant near-linear-dependence issues. In codes that eliminate `near-singular' eigenvectors of the overlap matrix (i.e., those for which the eigenvalue drops below a threshold), adding superfluous basis functions in general — and diffuse functions where they are unneeded in particular - can cause discontinuities on a correlated potential energy surface as orbitals drop in and out of the virtual space. When carrying out (semi)numerical frequency calculations, this can cause erratic results, as we observed here for acetylene.

In summary, we recommend the following:
\begin{itemize}
    \item If an accuracy of 20-30 cm$^{-1}$ is sufficient, or if anharmonicity's deviation from a simple scaling factor exceeds that level (and an anharmonic force field is not a practical option), then consider a DFT option such as $\omega$B97M-V\cite{Mardirossian2016} or the even more economical B97M-V\cite{Mardirossian2015}.
    \item If 10 cm$^{-1}$ is satisfactory, an empirical double hybrid like DSD-PBEP86 or revDSD-PBEP86 may be the right choice.
    \item For 4-5 cm$^{-1}$ accuracy, consider present G4-type approaches including G4-T$_{ano}$-v2 and G4-T$_{ano}$-F12-v1. Both methods share the same high-level correction [CCSD(T)-MP2]/ano-pVTZ, but G4-T$_{ano}$-v2 is combined with MP2/ano-pV5Z and G4-T$_{ano}$-F12-v1 uses MP2-F12/VTZ-F12. These two composite schemes offer similar accuracy and computational cost, and are suitable for larger molecules, particularly if analytic second derivatives at MP2 and CCSD(T) are available. It is worth noting that no empirical scaling parameters were employed in these top-performing approaches.
    \item For higher accuracy within the range of 1-2 cm$^{-1}$, it is important to extend beyond CCSD(T) as well as consider relativistic effects and the impact of diagonal Born-Oppenheimer corrections, especially in the case of hydrides.
\end{itemize}

\section{ACKNOWLEDGMENTS}
Research at Weizmann was supported by the Israel Science Foundation (grant 1969/20), by the Minerva Foundation (grant 2020/05), and 
by a research grant from the Artificial Intelligence for Smart Materials Research Fund, in Memory of Dr. Uriel Arnon. The authors would like to thank Dr. Mark Vilensky for technical assistance with ChemFarm, and Dr. Margarita Shepelenko for critical reading of the manuscript prior to submission.
ES thanks the Feinberg Graduate School (Weizmann Institute of Science) for a doctoral fellowship and the Onassis Foundation (Scholarship ID: FZP 052-2/2021-2022). \\

\section*{Supporting information and data availability statement}

A Microsoft Excel Workbook with all calculated harmonic frequencies for the HFREQ2014 dataset and the benzene molecule can be found at the DOI \url{http://doi.org/10.34933/2b615f91-8187-47c1-a29b-86705932634a}, or the shortened URL \url{http://tinyurl.com/G4compfreq}~. The \texttt{hfreq} implementation for geometry optimization and calculation of harmonic vibrational frequencies with composite energy schemes is available on GitHub at \url{http://github.com/msemidalas/hfreq}~. Additional data can be obtained from the authors upon reasonable request.

\bibliographystyle{achemso}%
\bibliography{Composite_WFN_vibfreq}%

\providecommand{\latin}[1]{#1}
\makeatletter
\providecommand{\doi}
  {\begingroup\let\do\@makeother\dospecials
  \catcode`\{=1 \catcode`\}=2 \doi@aux}
\providecommand{\doi@aux}[1]{\endgroup\texttt{#1}}
\makeatother
\providecommand*\mcitethebibliography{\thebibliography}
\csname @ifundefined\endcsname{endmcitethebibliography}
  {\let\endmcitethebibliography\endthebibliography}{}
\begin{mcitethebibliography}{152}
\providecommand*\natexlab[1]{#1}
\providecommand*\mciteSetBstSublistMode[1]{}
\providecommand*\mciteSetBstMaxWidthForm[2]{}
\providecommand*\mciteBstWouldAddEndPuncttrue
  {\def\EndOfBibitem{\unskip.}}
\providecommand*\mciteBstWouldAddEndPunctfalse
  {\let\EndOfBibitem\relax}
\providecommand*\mciteSetBstMidEndSepPunct[3]{}
\providecommand*\mciteSetBstSublistLabelBeginEnd[3]{}
\providecommand*\EndOfBibitem{}
\mciteSetBstSublistMode{f}
\mciteSetBstMaxWidthForm{subitem}{(\alph{mcitesubitemcount})}
\mciteSetBstSublistLabelBeginEnd
  {\mcitemaxwidthsubitemform\space}
  {\relax}
  {\relax}

\bibitem[Yurchenko(2023)]{Yurchenko2023}
Yurchenko,~S. \emph{{Computational Spectroscopy of Polyatomic Molecules}}; CRC
  Press: Boca Raton, 2023\relax
\mciteBstWouldAddEndPuncttrue
\mciteSetBstMidEndSepPunct{\mcitedefaultmidpunct}
{\mcitedefaultendpunct}{\mcitedefaultseppunct}\relax
\EndOfBibitem
\bibitem[Puzzarini \latin{et~al.}(2010)Puzzarini, Stanton, and
  Gauss]{Puzzarini2010}
Puzzarini,~C.; Stanton,~J.~F.; Gauss,~J. {Quantum-chemical calculation of
  spectroscopic parameters for rotational spectroscopy}. \emph{International
  Reviews in Physical Chemistry} \textbf{2010}, \emph{29}, 273--367\relax
\mciteBstWouldAddEndPuncttrue
\mciteSetBstMidEndSepPunct{\mcitedefaultmidpunct}
{\mcitedefaultendpunct}{\mcitedefaultseppunct}\relax
\EndOfBibitem
\bibitem[Puzzarini \latin{et~al.}(2019)Puzzarini, Bloino, Tasinato, and
  Barone]{Puzzarini2019}
Puzzarini,~C.; Bloino,~J.; Tasinato,~N.; Barone,~V. {Accuracy and
  Interpretability: The Devil and the Holy Grail. New Routes across Old
  Boundaries in Computational Spectroscopy}. \emph{Chem. Rev.} \textbf{2019},
  \emph{119}, 8131--8191\relax
\mciteBstWouldAddEndPuncttrue
\mciteSetBstMidEndSepPunct{\mcitedefaultmidpunct}
{\mcitedefaultendpunct}{\mcitedefaultseppunct}\relax
\EndOfBibitem
\bibitem[Barone \latin{et~al.}(2021)Barone, Alessandrini, Biczysko, Cheeseman,
  Clary, McCoy, DiRisio, Neese, Melosso, and Puzzarini]{Barone2021}
Barone,~V.; Alessandrini,~S.; Biczysko,~M.; Cheeseman,~J.~R.; Clary,~D.~C.;
  McCoy,~A.~B.; DiRisio,~R.~J.; Neese,~F.; Melosso,~M.; Puzzarini,~C.
  {Computational molecular spectroscopy}. \emph{Nature Reviews Methods Primers}
  \textbf{2021}, \emph{1}, 38\relax
\mciteBstWouldAddEndPuncttrue
\mciteSetBstMidEndSepPunct{\mcitedefaultmidpunct}
{\mcitedefaultendpunct}{\mcitedefaultseppunct}\relax
\EndOfBibitem
\bibitem[Baiz \latin{et~al.}(2020)Baiz, B{\l}asiak, Bredenbeck, Cho, Choi,
  Corcelli, Dijkstra, Feng, Garrett-Roe, Ge, Hanson-Heine, Hirst, Jansen, Kwac,
  Kubarych, Londergan, Maekawa, Reppert, Saito, Roy, Skinner, Stock, Straub,
  Thielges, Tominaga, Tokmakoff, Torii, Wang, Webb, and Zanni]{Baiz2020}
Baiz,~C.~R. \latin{et~al.}  {Vibrational Spectroscopic Map, Vibrational
  Spectroscopy, and Intermolecular Interaction}. \emph{Chem. Rev.}
  \textbf{2020}, \emph{120}, 7152--7218\relax
\mciteBstWouldAddEndPuncttrue
\mciteSetBstMidEndSepPunct{\mcitedefaultmidpunct}
{\mcitedefaultendpunct}{\mcitedefaultseppunct}\relax
\EndOfBibitem
\bibitem[Steinhauser and Hiermaier(2009)Steinhauser, and
  Hiermaier]{Steinhauser2009}
Steinhauser,~M.; Hiermaier,~S. {A Review of Computational Methods in Materials
  Science: Examples from Shock-Wave and Polymer Physics}. \emph{International
  Journal of Molecular Sciences} \textbf{2009}, \emph{10}, 5135--5216\relax
\mciteBstWouldAddEndPuncttrue
\mciteSetBstMidEndSepPunct{\mcitedefaultmidpunct}
{\mcitedefaultendpunct}{\mcitedefaultseppunct}\relax
\EndOfBibitem
\bibitem[Paul and Birol(2019)Paul, and Birol]{Paul2019}
Paul,~A.; Birol,~T. {Applications of DFT + DMFT in Materials Science}.
  \emph{Annual Review of Materials Research} \textbf{2019}, \emph{49},
  31--52\relax
\mciteBstWouldAddEndPuncttrue
\mciteSetBstMidEndSepPunct{\mcitedefaultmidpunct}
{\mcitedefaultendpunct}{\mcitedefaultseppunct}\relax
\EndOfBibitem
\bibitem[Stanton and Gauss(2003)Stanton, and Gauss]{Stanton2003}
Stanton,~J.~F.; Gauss,~J. \emph{Adv. Chem. Phys.}; 2003; Vol. 125; pp
  101--146\relax
\mciteBstWouldAddEndPuncttrue
\mciteSetBstMidEndSepPunct{\mcitedefaultmidpunct}
{\mcitedefaultendpunct}{\mcitedefaultseppunct}\relax
\EndOfBibitem
\bibitem[Karton and Martin(2010)Karton, and Martin]{jmlm230}
Karton,~A.; Martin,~J.~M. {Performance of W4 theory for spectroscopic constants
  and electrical properties of small molecules}. \emph{J. Chem. Phys.}
  \textbf{2010}, \emph{133}, 1--17\relax
\mciteBstWouldAddEndPuncttrue
\mciteSetBstMidEndSepPunct{\mcitedefaultmidpunct}
{\mcitedefaultendpunct}{\mcitedefaultseppunct}\relax
\EndOfBibitem
\bibitem[Raghavachari \latin{et~al.}(1989)Raghavachari, Trucks, Pople, and
  Head-Gordon]{Raghavachari1989}
Raghavachari,~K.; Trucks,~G.~W.; Pople,~J.~A.; Head-Gordon,~M. {A fifth-order
  perturbation comparison of electron correlation theories}. \emph{Chem. Phys.
  Lett.} \textbf{1989}, \emph{157}, 479--483\relax
\mciteBstWouldAddEndPuncttrue
\mciteSetBstMidEndSepPunct{\mcitedefaultmidpunct}
{\mcitedefaultendpunct}{\mcitedefaultseppunct}\relax
\EndOfBibitem
\bibitem[K{\'{a}}llay and Gauss(2005)K{\'{a}}llay, and Gauss]{Kallay2005}
K{\'{a}}llay,~M.; Gauss,~J. {Approximate treatment of higher excitations in
  coupled-cluster theory}. \emph{J. Chem. Phys.} \textbf{2005}, \emph{123},
  214105\relax
\mciteBstWouldAddEndPuncttrue
\mciteSetBstMidEndSepPunct{\mcitedefaultmidpunct}
{\mcitedefaultendpunct}{\mcitedefaultseppunct}\relax
\EndOfBibitem
\bibitem[Schwartz(1962)]{Schwartz1962a}
Schwartz,~C. {Importance of Angular Correlations between Atomic Electrons}.
  \emph{Phys. Rev.} \textbf{1962}, \emph{126}, 1015--1019\relax
\mciteBstWouldAddEndPuncttrue
\mciteSetBstMidEndSepPunct{\mcitedefaultmidpunct}
{\mcitedefaultendpunct}{\mcitedefaultseppunct}\relax
\EndOfBibitem
\bibitem[Hill(1985)]{Hill1985}
Hill,~R.~N. {Rates of convergence and error estimation formulas for the
  Rayleigh–Ritz variational method}. \emph{J. Chem. Phys.} \textbf{1985},
  \emph{83}, 1173--1196\relax
\mciteBstWouldAddEndPuncttrue
\mciteSetBstMidEndSepPunct{\mcitedefaultmidpunct}
{\mcitedefaultendpunct}{\mcitedefaultseppunct}\relax
\EndOfBibitem
\bibitem[Kutzelnigg and Morgan(1992)Kutzelnigg, and Morgan]{Kutzelnigg1992a}
Kutzelnigg,~W.; Morgan,~J.~D. {Rates of convergence of the partial‐wave
  expansions of atomic correlation energies}. \emph{J. Chem. Phys.}
  \textbf{1992}, \emph{96}, 4484--4508\relax
\mciteBstWouldAddEndPuncttrue
\mciteSetBstMidEndSepPunct{\mcitedefaultmidpunct}
{\mcitedefaultendpunct}{\mcitedefaultseppunct}\relax
\EndOfBibitem
\bibitem[Karton(2022)]{Karton2022}
Karton,~A. \emph{Annual Reports in Computational Chemistry}, 1st ed.; Elsevier
  B.V., 2022; Vol.~18; pp 123--166\relax
\mciteBstWouldAddEndPuncttrue
\mciteSetBstMidEndSepPunct{\mcitedefaultmidpunct}
{\mcitedefaultendpunct}{\mcitedefaultseppunct}\relax
\EndOfBibitem
\bibitem[Pople \latin{et~al.}(1989)Pople, Head‐Gordon, Fox, Raghavachari, and
  Curtiss]{Pople1989}
Pople,~J.~A.; Head‐Gordon,~M.; Fox,~D.~J.; Raghavachari,~K.; Curtiss,~L.~A.
  {Gaussian‐1 theory: A general procedure for prediction of molecular
  energies}. \emph{J. Chem. Phys.} \textbf{1989}, \emph{90}, 5622--5629\relax
\mciteBstWouldAddEndPuncttrue
\mciteSetBstMidEndSepPunct{\mcitedefaultmidpunct}
{\mcitedefaultendpunct}{\mcitedefaultseppunct}\relax
\EndOfBibitem
\bibitem[Curtiss \latin{et~al.}(1991)Curtiss, Raghavachari, Trucks, and
  Pople]{Curtiss1991}
Curtiss,~L.~A.; Raghavachari,~K.; Trucks,~G.~W.; Pople,~J.~A. {Gaussian‐2
  theory for molecular energies of first‐ and second‐row compounds}.
  \emph{J. Chem. Phys.} \textbf{1991}, \emph{94}, 7221--7230\relax
\mciteBstWouldAddEndPuncttrue
\mciteSetBstMidEndSepPunct{\mcitedefaultmidpunct}
{\mcitedefaultendpunct}{\mcitedefaultseppunct}\relax
\EndOfBibitem
\bibitem[Curtiss \latin{et~al.}(1995)Curtiss, McGrath, Blaudeau, Davis,
  Binning, and Radom]{Curtiss1995}
Curtiss,~L.~A.; McGrath,~M.~P.; Blaudeau,~J.-P.; Davis,~N.~E.; Binning,~R.~C.;
  Radom,~L. {Extension of Gaussian‐2 theory to molecules containing
  third‐row atoms Ga–Kr}. \emph{J. Chem. Phys.} \textbf{1995}, \emph{103},
  6104--6113\relax
\mciteBstWouldAddEndPuncttrue
\mciteSetBstMidEndSepPunct{\mcitedefaultmidpunct}
{\mcitedefaultendpunct}{\mcitedefaultseppunct}\relax
\EndOfBibitem
\bibitem[Baboul \latin{et~al.}(1999)Baboul, Curtiss, Redfern, and
  Raghavachari]{Baboul1999}
Baboul,~A.~G.; Curtiss,~L.~A.; Redfern,~P.~C.; Raghavachari,~K. {Gaussian-3
  theory using density functional geometries and zero-point energies}. \emph{J.
  Chem. Phys.} \textbf{1999}, \emph{110}, 7650--7657\relax
\mciteBstWouldAddEndPuncttrue
\mciteSetBstMidEndSepPunct{\mcitedefaultmidpunct}
{\mcitedefaultendpunct}{\mcitedefaultseppunct}\relax
\EndOfBibitem
\bibitem[Curtiss \latin{et~al.}(2007)Curtiss, Redfern, and
  Raghavachari]{Curtiss2007}
Curtiss,~L.~A.; Redfern,~P.~C.; Raghavachari,~K. {Gaussian-4 theory using
  reduced order perturbation theory}. \emph{J. Chem. Phys.} \textbf{2007},
  \emph{127}, 124105\relax
\mciteBstWouldAddEndPuncttrue
\mciteSetBstMidEndSepPunct{\mcitedefaultmidpunct}
{\mcitedefaultendpunct}{\mcitedefaultseppunct}\relax
\EndOfBibitem
\bibitem[Curtiss \latin{et~al.}(2007)Curtiss, Redfern, and
  Raghavachari]{Curtiss2007a}
Curtiss,~L.~A.; Redfern,~P.~C.; Raghavachari,~K. {Gaussian-4 theory}. \emph{J.
  Chem. Phys.} \textbf{2007}, \emph{126}, 084108\relax
\mciteBstWouldAddEndPuncttrue
\mciteSetBstMidEndSepPunct{\mcitedefaultmidpunct}
{\mcitedefaultendpunct}{\mcitedefaultseppunct}\relax
\EndOfBibitem
\bibitem[Montgomery \latin{et~al.}(1999)Montgomery, Frisch, Ochterski, and
  Petersson]{Montgomery1999}
Montgomery,~J.~A.; Frisch,~M.~J.; Ochterski,~J.~W.; Petersson,~G.~A. {A
  complete basis set model chemistry. VI. Use of density functional geometries
  and frequencies}. \emph{J. Chem. Phys.} \textbf{1999}, \emph{110},
  2822--2827\relax
\mciteBstWouldAddEndPuncttrue
\mciteSetBstMidEndSepPunct{\mcitedefaultmidpunct}
{\mcitedefaultendpunct}{\mcitedefaultseppunct}\relax
\EndOfBibitem
\bibitem[Montgomery \latin{et~al.}(2000)Montgomery, Frisch, Ochterski, and
  Petersson]{Montgomery2000}
Montgomery,~J.~A.; Frisch,~M.~J.; Ochterski,~J.~W.; Petersson,~G.~A. {A
  complete basis set model chemistry. VII. Use of the minimum population
  localization method}. \emph{J. Chem. Phys.} \textbf{2000}, \emph{112},
  6532--6542\relax
\mciteBstWouldAddEndPuncttrue
\mciteSetBstMidEndSepPunct{\mcitedefaultmidpunct}
{\mcitedefaultendpunct}{\mcitedefaultseppunct}\relax
\EndOfBibitem
\bibitem[Petersson(2001)]{Petersson2001}
Petersson,~G.~A. In \emph{Quantum-Mechanical Prediction of Thermochemical
  Data}; Cioslowski,~J., Ed.; Kluwer Academic Publishers: Dordrecht, 2001; pp
  99--130\relax
\mciteBstWouldAddEndPuncttrue
\mciteSetBstMidEndSepPunct{\mcitedefaultmidpunct}
{\mcitedefaultendpunct}{\mcitedefaultseppunct}\relax
\EndOfBibitem
\bibitem[Martin and de~Oliveira(1999)Martin, and de~Oliveira]{Martin1999}
Martin,~J. M.~L.; de~Oliveira,~G. {Towards standard methods for benchmark
  quality {\em ab initio} thermochemistry—W1 and W2 theory}. \emph{J. Chem.
  Phys.} \textbf{1999}, \emph{111}, 1843--1856\relax
\mciteBstWouldAddEndPuncttrue
\mciteSetBstMidEndSepPunct{\mcitedefaultmidpunct}
{\mcitedefaultendpunct}{\mcitedefaultseppunct}\relax
\EndOfBibitem
\bibitem[Parthiban and Martin(2001)Parthiban, and Martin]{jmlm148}
Parthiban,~S.; Martin,~J. M.~L. {Assessment of W1 and W2 theories for the
  computation of electron affinities, ionization potentials, heats of
  formation, and proton affinities}. \emph{J. Chem. Phys.} \textbf{2001},
  \emph{114}, 6014--6029\relax
\mciteBstWouldAddEndPuncttrue
\mciteSetBstMidEndSepPunct{\mcitedefaultmidpunct}
{\mcitedefaultendpunct}{\mcitedefaultseppunct}\relax
\EndOfBibitem
\bibitem[Martin and Parthiban(2002)Martin, and Parthiban]{jmlm151}
Martin,~J. M.~L.; Parthiban,~S. In \emph{Quantum-Mechanical Predict.
  Thermochem. Data}; Cioslowski,~J., Ed.; Understanding Chemical Reactivity;
  Kluwer Academic Publishers: Dordrecht, 2002; Vol.~22; pp 31--65\relax
\mciteBstWouldAddEndPuncttrue
\mciteSetBstMidEndSepPunct{\mcitedefaultmidpunct}
{\mcitedefaultendpunct}{\mcitedefaultseppunct}\relax
\EndOfBibitem
\bibitem[DeYonker \latin{et~al.}(2006)DeYonker, Cundari, and
  Wilson]{DeYonker2006}
DeYonker,~N.~J.; Cundari,~T.~R.; Wilson,~A.~K. {The correlation consistent
  composite approach (ccCA): An alternative to the Gaussian-n methods}.
  \emph{J. Chem. Phys.} \textbf{2006}, \emph{124}, 114104\relax
\mciteBstWouldAddEndPuncttrue
\mciteSetBstMidEndSepPunct{\mcitedefaultmidpunct}
{\mcitedefaultendpunct}{\mcitedefaultseppunct}\relax
\EndOfBibitem
\bibitem[DeYonker \latin{et~al.}(2009)DeYonker, Wilson, Pierpont, Cundari, and
  Wilson]{DeYonker2009}
DeYonker,~N.~J.; Wilson,~B.~R.; Pierpont,~A.~W.; Cundari,~T.~R.; Wilson,~A.~K.
  {Towards the intrinsic error of the correlation consistent Composite Approach
  (ccCA)}. \emph{Mol. Phys.} \textbf{2009}, \emph{107}, 1107--1121\relax
\mciteBstWouldAddEndPuncttrue
\mciteSetBstMidEndSepPunct{\mcitedefaultmidpunct}
{\mcitedefaultendpunct}{\mcitedefaultseppunct}\relax
\EndOfBibitem
\bibitem[Peterson \latin{et~al.}(2016)Peterson, Penchoff, and
  Wilson]{Peterson2016}
Peterson,~C.; Penchoff,~D.; Wilson,~A. \emph{Annual Reports in Computational
  Chemistry}; Elsevier, 2016; Vol.~12; pp 3--45\relax
\mciteBstWouldAddEndPuncttrue
\mciteSetBstMidEndSepPunct{\mcitedefaultmidpunct}
{\mcitedefaultendpunct}{\mcitedefaultseppunct}\relax
\EndOfBibitem
\bibitem[Curtiss \latin{et~al.}(1993)Curtiss, Raghavachari, and
  Pople]{Curtiss1993}
Curtiss,~L.~A.; Raghavachari,~K.; Pople,~J.~A. {Gaussian-2 theory using reduced
  M{\o}ller-Plesset orders}. \emph{J. Chem. Phys.} \textbf{1993}, \emph{98},
  1293--1298\relax
\mciteBstWouldAddEndPuncttrue
\mciteSetBstMidEndSepPunct{\mcitedefaultmidpunct}
{\mcitedefaultendpunct}{\mcitedefaultseppunct}\relax
\EndOfBibitem
\bibitem[Curtiss \latin{et~al.}(1999)Curtiss, Redfern, Raghavachari, Rassolov,
  and Pople]{Curtiss1999}
Curtiss,~L.~A.; Redfern,~P.~C.; Raghavachari,~K.; Rassolov,~V.; Pople,~J.~A.
  {Gaussian-3 theory using reduced M{\o}ller-Plesset order}. \emph{J. Chem.
  Phys.} \textbf{1999}, \emph{110}, 4703--4709\relax
\mciteBstWouldAddEndPuncttrue
\mciteSetBstMidEndSepPunct{\mcitedefaultmidpunct}
{\mcitedefaultendpunct}{\mcitedefaultseppunct}\relax
\EndOfBibitem
\bibitem[Chan \latin{et~al.}(2011)Chan, Deng, and Radom]{Chan2011}
Chan,~B.; Deng,~J.; Radom,~L. {G4(MP2)-6X: A Cost-Effective Improvement to
  G4(MP2)}. \emph{J. Chem. Theor. Comput.} \textbf{2011}, \emph{7},
  112--120\relax
\mciteBstWouldAddEndPuncttrue
\mciteSetBstMidEndSepPunct{\mcitedefaultmidpunct}
{\mcitedefaultendpunct}{\mcitedefaultseppunct}\relax
\EndOfBibitem
\bibitem[Chan \latin{et~al.}(2019)Chan, Karton, and Raghavachari]{Chan2019b}
Chan,~B.; Karton,~A.; Raghavachari,~K. {G4(MP2)-XK: A Variant of the G4(MP2)-6X
  Composite Method with Expanded Applicability for Main-Group Elements up to
  Radon}. \emph{J. Chem. Theor. Comput.} \textbf{2019}, \emph{15},
  4478--4484\relax
\mciteBstWouldAddEndPuncttrue
\mciteSetBstMidEndSepPunct{\mcitedefaultmidpunct}
{\mcitedefaultendpunct}{\mcitedefaultseppunct}\relax
\EndOfBibitem
\bibitem[Semidalas and Martin(2020)Semidalas, and Martin]{Semidalas2020a}
Semidalas,~E.; Martin,~J. M.~L. {Canonical and DLPNO-Based G4(MP2)XK-Inspired
  Composite Wave Function Methods Parametrized against Large and Chemically
  Diverse Training Sets: Are They More Accurate and/or Robust than
  Double-Hybrid DFT?} \emph{J. Chem. Theor. Comput.} \textbf{2020}, \emph{16},
  4238--4255\relax
\mciteBstWouldAddEndPuncttrue
\mciteSetBstMidEndSepPunct{\mcitedefaultmidpunct}
{\mcitedefaultendpunct}{\mcitedefaultseppunct}\relax
\EndOfBibitem
\bibitem[Semidalas and Martin(2020)Semidalas, and Martin]{Semidalas2020}
Semidalas,~E.; Martin,~J. M.~L. {Canonical and DLPNO-Based Composite
  Wavefunction Methods Parametrized against Large and Chemically Diverse
  Training Sets. 2: Correlation-Consistent Basis Sets, Core–Valence
  Correlation, and F12 Alternatives}. \emph{J. Chem. Theor. Comput.}
  \textbf{2020}, \emph{16}, 7507--7524\relax
\mciteBstWouldAddEndPuncttrue
\mciteSetBstMidEndSepPunct{\mcitedefaultmidpunct}
{\mcitedefaultendpunct}{\mcitedefaultseppunct}\relax
\EndOfBibitem
\bibitem[Goerigk \latin{et~al.}(2017)Goerigk, Hansen, Bauer, Ehrlich, Najibi,
  and Grimme]{Goerigk2017a}
Goerigk,~L.; Hansen,~A.; Bauer,~C.; Ehrlich,~S.; Najibi,~A.; Grimme,~S. {A look
  at the density functional theory zoo with the advanced GMTKN55 database for
  general main group thermochemistry, kinetics and noncovalent interactions}.
  \emph{Phys. Chem. Chem. Phys.} \textbf{2017}, \emph{19}, 32184--32215\relax
\mciteBstWouldAddEndPuncttrue
\mciteSetBstMidEndSepPunct{\mcitedefaultmidpunct}
{\mcitedefaultendpunct}{\mcitedefaultseppunct}\relax
\EndOfBibitem
\bibitem[Ten-no and Noga(2012)Ten-no, and Noga]{Ten-no2012}
Ten-no,~S.; Noga,~J. {Explicitly correlated electronic structure theory from
  R12/F12 ans{\"{a}}tze}. \emph{WIREs Computational Molecular Science}
  \textbf{2012}, \emph{2}, 114--125\relax
\mciteBstWouldAddEndPuncttrue
\mciteSetBstMidEndSepPunct{\mcitedefaultmidpunct}
{\mcitedefaultendpunct}{\mcitedefaultseppunct}\relax
\EndOfBibitem
\bibitem[Ten-no(2012)]{Ten-no2012a}
Ten-no,~S. {Explicitly correlated wave functions: summary and perspective}.
  \emph{Theoretical Chemistry Accounts} \textbf{2012}, \emph{131}, 1070\relax
\mciteBstWouldAddEndPuncttrue
\mciteSetBstMidEndSepPunct{\mcitedefaultmidpunct}
{\mcitedefaultendpunct}{\mcitedefaultseppunct}\relax
\EndOfBibitem
\bibitem[Kong \latin{et~al.}(2012)Kong, Bischoff, and Valeev]{Kong2012}
Kong,~L.; Bischoff,~F.~A.; Valeev,~E.~F. {Explicitly Correlated R12/F12 Methods
  for Electronic Structure}. \emph{Chem. Rev.} \textbf{2012}, \emph{112},
  75--107\relax
\mciteBstWouldAddEndPuncttrue
\mciteSetBstMidEndSepPunct{\mcitedefaultmidpunct}
{\mcitedefaultendpunct}{\mcitedefaultseppunct}\relax
\EndOfBibitem
\bibitem[H{\"{a}}ttig \latin{et~al.}(2012)H{\"{a}}ttig, Klopper, K{\"{o}}hn,
  and Tew]{Hattig2012a}
H{\"{a}}ttig,~C.; Klopper,~W.; K{\"{o}}hn,~A.; Tew,~D.~P. {Explicitly
  Correlated Electrons in Molecules}. \emph{Chem. Rev.} \textbf{2012},
  \emph{112}, 4--74\relax
\mciteBstWouldAddEndPuncttrue
\mciteSetBstMidEndSepPunct{\mcitedefaultmidpunct}
{\mcitedefaultendpunct}{\mcitedefaultseppunct}\relax
\EndOfBibitem
\bibitem[Klopper and Kutzelnigg(1987)Klopper, and Kutzelnigg]{Klopper1987}
Klopper,~W.; Kutzelnigg,~W. {M{\o}ller-Plesset calculations taking care of the
  correlation cusp}. \emph{Chem. Phys. Lett.} \textbf{1987}, \emph{134},
  17--22\relax
\mciteBstWouldAddEndPuncttrue
\mciteSetBstMidEndSepPunct{\mcitedefaultmidpunct}
{\mcitedefaultendpunct}{\mcitedefaultseppunct}\relax
\EndOfBibitem
\bibitem[Kutzelnigg and Klopper(1991)Kutzelnigg, and Klopper]{Kutzelnigg1991}
Kutzelnigg,~W.; Klopper,~W. {Wave functions with terms linear in the
  interelectronic coordinates to take care of the correlation cusp. I. General
  theory}. \emph{J. Chem. Phys.} \textbf{1991}, \emph{94}, 1985--2001\relax
\mciteBstWouldAddEndPuncttrue
\mciteSetBstMidEndSepPunct{\mcitedefaultmidpunct}
{\mcitedefaultendpunct}{\mcitedefaultseppunct}\relax
\EndOfBibitem
\bibitem[Karton and Martin(2012)Karton, and Martin]{Karton2012}
Karton,~A.; Martin,~J. M.~L. {Explicitly correlated Wn theory: W1-F12 and
  W2-F12}. \emph{J. Chem. Phys.} \textbf{2012}, \emph{136}, 124114\relax
\mciteBstWouldAddEndPuncttrue
\mciteSetBstMidEndSepPunct{\mcitedefaultmidpunct}
{\mcitedefaultendpunct}{\mcitedefaultseppunct}\relax
\EndOfBibitem
\bibitem[Mahler and Wilson(2013)Mahler, and Wilson]{Mahler2013}
Mahler,~A.; Wilson,~A.~K. {Explicitly Correlated Methods within the ccCA
  Methodology}. \emph{J. Chem. Theor. Comput.} \textbf{2013}, \emph{9},
  1402--1407\relax
\mciteBstWouldAddEndPuncttrue
\mciteSetBstMidEndSepPunct{\mcitedefaultmidpunct}
{\mcitedefaultendpunct}{\mcitedefaultseppunct}\relax
\EndOfBibitem
\bibitem[Ventura \latin{et~al.}(2021)Ventura, Kieninger, Katz, Vega‐Teijido,
  Segovia, and Irving]{Ventura2021}
Ventura,~O.~N.; Kieninger,~M.; Katz,~A.; Vega‐Teijido,~M.; Segovia,~M.;
  Irving,~K. {SVECV‐f12: Benchmark of a composite scheme for accurate and
  cost effective evaluation of reaction barriers}. \emph{Int. J. Quantum Chem.}
  \textbf{2021}, \emph{121}, 1--19\relax
\mciteBstWouldAddEndPuncttrue
\mciteSetBstMidEndSepPunct{\mcitedefaultmidpunct}
{\mcitedefaultendpunct}{\mcitedefaultseppunct}\relax
\EndOfBibitem
\bibitem[Stanton(1997)]{Stanton1997}
Stanton,~J.~F. {Why CCSD(T) works: A different perspective}. \emph{Chem. Phys.
  Lett.} \textbf{1997}, \emph{281}, 130--134\relax
\mciteBstWouldAddEndPuncttrue
\mciteSetBstMidEndSepPunct{\mcitedefaultmidpunct}
{\mcitedefaultendpunct}{\mcitedefaultseppunct}\relax
\EndOfBibitem
\bibitem[Tajti \latin{et~al.}(2004)Tajti, Szalay, Cs{\'{a}}sz{\'{a}}r,
  K{\'{a}}llay, Gauss, Valeev, Flowers, V{\'{a}}zquez, and Stanton]{Tajti2004}
Tajti,~A.; Szalay,~P.~G.; Cs{\'{a}}sz{\'{a}}r,~A.~G.; K{\'{a}}llay,~M.;
  Gauss,~J.; Valeev,~E.~F.; Flowers,~B.~A.; V{\'{a}}zquez,~J.; Stanton,~J.~F.
  {HEAT: High accuracy extrapolated {\em ab initio} thermochemistry}. \emph{J.
  Chem. Phys.} \textbf{2004}, \emph{121}, 11599--11613\relax
\mciteBstWouldAddEndPuncttrue
\mciteSetBstMidEndSepPunct{\mcitedefaultmidpunct}
{\mcitedefaultendpunct}{\mcitedefaultseppunct}\relax
\EndOfBibitem
\bibitem[Bomble \latin{et~al.}(2006)Bomble, V{\'{a}}zquez, K{\'{a}}llay,
  Michauk, Szalay, Cs{\'{a}}sz{\'{a}}r, Gauss, and Stanton]{Bomble2006}
Bomble,~Y.~J.; V{\'{a}}zquez,~J.; K{\'{a}}llay,~M.; Michauk,~C.; Szalay,~P.~G.;
  Cs{\'{a}}sz{\'{a}}r,~A.~G.; Gauss,~J.; Stanton,~J.~F. {High-accuracy
  extrapolated {\em ab initio} thermochemistry. II. Minor improvements to the
  protocol and a vital simplification}. \emph{J. Chem. Phys.} \textbf{2006},
  \emph{125}, 064108\relax
\mciteBstWouldAddEndPuncttrue
\mciteSetBstMidEndSepPunct{\mcitedefaultmidpunct}
{\mcitedefaultendpunct}{\mcitedefaultseppunct}\relax
\EndOfBibitem
\bibitem[Harding \latin{et~al.}(2008)Harding, V{\'{a}}zquez, Ruscic, Wilson,
  Gauss, and Stanton]{Harding2008}
Harding,~M.~E.; V{\'{a}}zquez,~J.; Ruscic,~B.; Wilson,~A.~K.; Gauss,~J.;
  Stanton,~J.~F. {High-accuracy extrapolated {\em ab initio} thermochemistry.
  III. Additional improvements and overview}. \emph{J. Chem. Phys.}
  \textbf{2008}, \emph{128}, 114111\relax
\mciteBstWouldAddEndPuncttrue
\mciteSetBstMidEndSepPunct{\mcitedefaultmidpunct}
{\mcitedefaultendpunct}{\mcitedefaultseppunct}\relax
\EndOfBibitem
\bibitem[Thorpe \latin{et~al.}(2019)Thorpe, Lopez, Nguyen, Baraban, Bross,
  Ruscic, and Stanton]{Thorpe2019}
Thorpe,~J.~H.; Lopez,~C.~A.; Nguyen,~T.~L.; Baraban,~J.~H.; Bross,~D.~H.;
  Ruscic,~B.; Stanton,~J.~F. {High-accuracy extrapolated {\em ab initio}
  thermochemistry. IV. A modified recipe for computational efficiency}.
  \emph{J. Chem. Phys.} \textbf{2019}, \emph{150}, 224102\relax
\mciteBstWouldAddEndPuncttrue
\mciteSetBstMidEndSepPunct{\mcitedefaultmidpunct}
{\mcitedefaultendpunct}{\mcitedefaultseppunct}\relax
\EndOfBibitem
\bibitem[Dixon \latin{et~al.}(2012)Dixon, Feller, and Peterson]{Dixon2012}
Dixon,~D.~A.; Feller,~D.; Peterson,~K.~A. \emph{Annual Reports in Computational
  Chemistry}; Elsevier, 2012; Vol.~8; pp 1--28\relax
\mciteBstWouldAddEndPuncttrue
\mciteSetBstMidEndSepPunct{\mcitedefaultmidpunct}
{\mcitedefaultendpunct}{\mcitedefaultseppunct}\relax
\EndOfBibitem
\bibitem[Feller \latin{et~al.}(2014)Feller, Peterson, and Ruscic]{Feller2014}
Feller,~D.; Peterson,~K.~A.; Ruscic,~B. {Improved accuracy benchmarks of small
  molecules using correlation consistent basis sets}. \emph{Theoretical
  Chemistry Accounts} \textbf{2014}, \emph{133}, 1407\relax
\mciteBstWouldAddEndPuncttrue
\mciteSetBstMidEndSepPunct{\mcitedefaultmidpunct}
{\mcitedefaultendpunct}{\mcitedefaultseppunct}\relax
\EndOfBibitem
\bibitem[Feller \latin{et~al.}(2016)Feller, Peterson, and Dixon]{Feller2016}
Feller,~D.; Peterson,~K.; Dixon,~D. \emph{Annual Reports in Computational
  Chemistry}; Elsevier, 2016; Vol.~12; pp 47--78\relax
\mciteBstWouldAddEndPuncttrue
\mciteSetBstMidEndSepPunct{\mcitedefaultmidpunct}
{\mcitedefaultendpunct}{\mcitedefaultseppunct}\relax
\EndOfBibitem
\bibitem[Boese \latin{et~al.}(2004)Boese, Oren, Atasoylu, Martin, K{\'{a}}llay,
  and Gauss]{jmlm173}
Boese,~A.~D.; Oren,~M.; Atasoylu,~O.; Martin,~J. M.~L.; K{\'{a}}llay,~M.;
  Gauss,~J. {W3 theory: Robust computational thermochemistry in the kJ/mol
  accuracy range}. \emph{J. Chem. Phys.} \textbf{2004}, \emph{120},
  4129--4141\relax
\mciteBstWouldAddEndPuncttrue
\mciteSetBstMidEndSepPunct{\mcitedefaultmidpunct}
{\mcitedefaultendpunct}{\mcitedefaultseppunct}\relax
\EndOfBibitem
\bibitem[Karton \latin{et~al.}(2006)Karton, Rabinovich, Martin, and
  Ruscic]{Karton2006}
Karton,~A.; Rabinovich,~E.; Martin,~J. M.~L.; Ruscic,~B. {W4 theory for
  computational thermochemistry: In pursuit of confident sub-kJ/mol
  predictions}. \emph{J. Chem. Phys.} \textbf{2006}, \emph{125}, 144108\relax
\mciteBstWouldAddEndPuncttrue
\mciteSetBstMidEndSepPunct{\mcitedefaultmidpunct}
{\mcitedefaultendpunct}{\mcitedefaultseppunct}\relax
\EndOfBibitem
\bibitem[Sylvetsky \latin{et~al.}(2016)Sylvetsky, Peterson, Karton, and
  Martin]{Sylvetsky2016}
Sylvetsky,~N.; Peterson,~K.~A.; Karton,~A.; Martin,~J. M.~L. {Toward a W4-F12
  approach: Can explicitly correlated and orbital-based {\em ab initio} CCSD(T)
  limits be reconciled?} \emph{J. Chem. Phys.} \textbf{2016}, \emph{144},
  214101\relax
\mciteBstWouldAddEndPuncttrue
\mciteSetBstMidEndSepPunct{\mcitedefaultmidpunct}
{\mcitedefaultendpunct}{\mcitedefaultseppunct}\relax
\EndOfBibitem
\bibitem[Spiegel \latin{et~al.}(2023)Spiegel, Semidalas, Martin, Bentley, and
  Stanton]{Spiegel2023}
Spiegel,~M.; Semidalas,~E.; Martin,~J. M.~L.; Bentley,~M.~R.; Stanton,~J.~F.
  {Post-CCSD(T) corrections to bond distances and vibrational frequencies: the
  power of $\Lambda$}. \emph{Molecular Physics} \textbf{2023}, e2252114\relax
\mciteBstWouldAddEndPuncttrue
\mciteSetBstMidEndSepPunct{\mcitedefaultmidpunct}
{\mcitedefaultendpunct}{\mcitedefaultseppunct}\relax
\EndOfBibitem
\bibitem[Heckert \latin{et~al.}(2005)Heckert, K{\'{a}}llay, and
  Gauss]{Heckert2005}
Heckert,~M.; K{\'{a}}llay,~M.; Gauss,~J. {Molecular equilibrium geometries
  based on coupled-cluster calculations including quadruple excitations}.
  \emph{Mol. Phys.} \textbf{2005}, \emph{103}, 2109--2115\relax
\mciteBstWouldAddEndPuncttrue
\mciteSetBstMidEndSepPunct{\mcitedefaultmidpunct}
{\mcitedefaultendpunct}{\mcitedefaultseppunct}\relax
\EndOfBibitem
\bibitem[Heckert \latin{et~al.}(2006)Heckert, K{\'{a}}llay, Tew, Klopper, and
  Gauss]{Heckert2006}
Heckert,~M.; K{\'{a}}llay,~M.; Tew,~D.~P.; Klopper,~W.; Gauss,~J. {Basis-set
  extrapolation techniques for the accurate calculation of molecular
  equilibrium geometries using coupled-cluster theory}. \emph{J. Chem. Phys.}
  \textbf{2006}, \emph{125}, 044108\relax
\mciteBstWouldAddEndPuncttrue
\mciteSetBstMidEndSepPunct{\mcitedefaultmidpunct}
{\mcitedefaultendpunct}{\mcitedefaultseppunct}\relax
\EndOfBibitem
\bibitem[Puzzarini \latin{et~al.}(2008)Puzzarini, Heckert, and
  Gauss]{Puzzarini2008}
Puzzarini,~C.; Heckert,~M.; Gauss,~J. {The accuracy of rotational constants
  predicted by high-level quantum-chemical calculations. I. molecules
  containing first-row atoms}. \emph{J. Chem. Phys.} \textbf{2008},
  \emph{128}\relax
\mciteBstWouldAddEndPuncttrue
\mciteSetBstMidEndSepPunct{\mcitedefaultmidpunct}
{\mcitedefaultendpunct}{\mcitedefaultseppunct}\relax
\EndOfBibitem
\bibitem[Ruden \latin{et~al.}(2004)Ruden, Helgaker, J{\o}rgensen, and
  Olsen]{Ruden2004}
Ruden,~T.~A.; Helgaker,~T.; J{\o}rgensen,~P.; Olsen,~J. {Coupled-cluster
  connected quadruples and quintuples corrections to the harmonic vibrational
  frequencies and equilibrium bond distances of HF, N$_2$, F$_2$, and CO}.
  \emph{J. Chem. Phys.} \textbf{2004}, \emph{121}, 5874--5884\relax
\mciteBstWouldAddEndPuncttrue
\mciteSetBstMidEndSepPunct{\mcitedefaultmidpunct}
{\mcitedefaultendpunct}{\mcitedefaultseppunct}\relax
\EndOfBibitem
\bibitem[Morgan \latin{et~al.}(2018)Morgan, Matthews, Ringholm, Agarwal, Gong,
  Ruud, Allen, Stanton, and Schaefer]{Morgan2018}
Morgan,~W.~J.; Matthews,~D.~A.; Ringholm,~M.; Agarwal,~J.; Gong,~J.~Z.;
  Ruud,~K.; Allen,~W.~D.; Stanton,~J.~F.; Schaefer,~H.~F. {Geometric Energy
  Derivatives at the Complete Basis Set Limit: Application to the Equilibrium
  Structure and Molecular Force Field of Formaldehyde}. \emph{J. Chem. Theor.
  Comput.} \textbf{2018}, \emph{14}, 1333--1350\relax
\mciteBstWouldAddEndPuncttrue
\mciteSetBstMidEndSepPunct{\mcitedefaultmidpunct}
{\mcitedefaultendpunct}{\mcitedefaultseppunct}\relax
\EndOfBibitem
\bibitem[Allen \latin{et~al.}(1993)Allen, East, and
  Cs{\'{a}}sz{\'{a}}r]{Allen1993}
Allen,~W.~D.; East,~A. L.~L.; Cs{\'{a}}sz{\'{a}}r,~A.~G. In \emph{Structures
  and Conformations of Non-Rigid Molecules}; Laane,~J., Dakkouri,~M., van~der
  Veken,~B., Oberhammer,~H., Eds.; Springer Netherlands: Dordrecht, 1993; pp
  343--373\relax
\mciteBstWouldAddEndPuncttrue
\mciteSetBstMidEndSepPunct{\mcitedefaultmidpunct}
{\mcitedefaultendpunct}{\mcitedefaultseppunct}\relax
\EndOfBibitem
\bibitem[East \latin{et~al.}(1993)East, Johnson, and Allen]{East1993}
East,~A. L.~L.; Johnson,~C.~S.; Allen,~W.~D. {Characterization of the X 1A'
  state of isocyanic acid}. \emph{J. Chem. Phys.} \textbf{1993}, \emph{98},
  1299--1328\relax
\mciteBstWouldAddEndPuncttrue
\mciteSetBstMidEndSepPunct{\mcitedefaultmidpunct}
{\mcitedefaultendpunct}{\mcitedefaultseppunct}\relax
\EndOfBibitem
\bibitem[East and Allen(1993)East, and Allen]{East1993a}
East,~A. L.~L.; Allen,~W.~D. {The heat of formation of NCO}. \emph{J. Chem.
  Phys.} \textbf{1993}, \emph{99}, 4638--4650\relax
\mciteBstWouldAddEndPuncttrue
\mciteSetBstMidEndSepPunct{\mcitedefaultmidpunct}
{\mcitedefaultendpunct}{\mcitedefaultseppunct}\relax
\EndOfBibitem
\bibitem[Cs{\'{a}}sz{\'{a}}r \latin{et~al.}(1998)Cs{\'{a}}sz{\'{a}}r, Allen,
  and Schaefer]{Csaszar1998}
Cs{\'{a}}sz{\'{a}}r,~A.~G.; Allen,~W.~D.; Schaefer,~H.~F. {In pursuit of the
  {\em ab initio} limit for conformational energy prototypes}. \emph{J. Chem.
  Phys.} \textbf{1998}, \emph{108}, 9751--9764\relax
\mciteBstWouldAddEndPuncttrue
\mciteSetBstMidEndSepPunct{\mcitedefaultmidpunct}
{\mcitedefaultendpunct}{\mcitedefaultseppunct}\relax
\EndOfBibitem
\bibitem[Zhu and Xu(2023)Zhu, and Xu]{Zhu2023}
Zhu,~Z.; Xu,~X. {Focal-Point Analysis to Achieve Accurate CCSD(T) Data Set
  References for Static Polarizabilities}. \emph{J. Chem. Theor. Comput.}
  \textbf{2023}, \emph{19}, 3112--3122\relax
\mciteBstWouldAddEndPuncttrue
\mciteSetBstMidEndSepPunct{\mcitedefaultmidpunct}
{\mcitedefaultendpunct}{\mcitedefaultseppunct}\relax
\EndOfBibitem
\bibitem[Huang and Lee(2008)Huang, and Lee]{Huang2008}
Huang,~X.; Lee,~T.~J. {A procedure for computing accurate {\em ab initio}
  quartic force fields: Application to HO$_2^+$ and H$_2$O}. \emph{J. Chem.
  Phys.} \textbf{2008}, \emph{129}\relax
\mciteBstWouldAddEndPuncttrue
\mciteSetBstMidEndSepPunct{\mcitedefaultmidpunct}
{\mcitedefaultendpunct}{\mcitedefaultseppunct}\relax
\EndOfBibitem
\bibitem[Fortenberry and Lee(2019)Fortenberry, and Lee]{Fortenberry2019}
Fortenberry,~R.~C.; Lee,~T.~J. \emph{Annu. Rep. Comput. Chem.}, 1st ed.;
  Elsevier B.V., 2019; Vol.~15; pp 173--202\relax
\mciteBstWouldAddEndPuncttrue
\mciteSetBstMidEndSepPunct{\mcitedefaultmidpunct}
{\mcitedefaultendpunct}{\mcitedefaultseppunct}\relax
\EndOfBibitem
\bibitem[Gardner \latin{et~al.}(2021)Gardner, Westbrook, Fortenberry, and
  Lee]{Gardner2021}
Gardner,~M.~B.; Westbrook,~B.~R.; Fortenberry,~R.~C.; Lee,~T.~J.
  {Highly-accurate quartic force fields for the prediction of anharmonic
  rotational constants and fundamental vibrational frequencies}.
  \emph{Spectrochim. Acta A} \textbf{2021}, \emph{248}, 119184\relax
\mciteBstWouldAddEndPuncttrue
\mciteSetBstMidEndSepPunct{\mcitedefaultmidpunct}
{\mcitedefaultendpunct}{\mcitedefaultseppunct}\relax
\EndOfBibitem
\bibitem[Boese and Martin(2004)Boese, and Martin]{jmlm172}
Boese,~A.~D.; Martin,~J. M.~L. {Vibrational Spectra of the Azabenzenes
  Revisited: Anharmonic Force Fields}. \emph{J. Phys. Chem. A} \textbf{2004},
  \emph{108}, 3085--3096\relax
\mciteBstWouldAddEndPuncttrue
\mciteSetBstMidEndSepPunct{\mcitedefaultmidpunct}
{\mcitedefaultendpunct}{\mcitedefaultseppunct}\relax
\EndOfBibitem
\bibitem[Fan \latin{et~al.}(2006)Fan, Ho, and Bettens]{Bettens2006}
Fan,~Y.; Ho,~J.; Bettens,~R.~P. {Approximating coupled cluster level
  vibrational frequencies with composite methods}. \emph{J. Phys. Chem. A}
  \textbf{2006}, \emph{110}, 2796--2800\relax
\mciteBstWouldAddEndPuncttrue
\mciteSetBstMidEndSepPunct{\mcitedefaultmidpunct}
{\mcitedefaultendpunct}{\mcitedefaultseppunct}\relax
\EndOfBibitem
\bibitem[Barone \latin{et~al.}(2014)Barone, Biczysko, Bloino, and
  Puzzarini]{Barone2014}
Barone,~V.; Biczysko,~M.; Bloino,~J.; Puzzarini,~C. {Accurate molecular
  structures and infrared spectra of trans-2,3- dideuterooxirane,
  methyloxirane, and trans-2,3-dimethyloxirane}. \emph{J. Chem. Phys.}
  \textbf{2014}, \emph{141}\relax
\mciteBstWouldAddEndPuncttrue
\mciteSetBstMidEndSepPunct{\mcitedefaultmidpunct}
{\mcitedefaultendpunct}{\mcitedefaultseppunct}\relax
\EndOfBibitem
\bibitem[Martin and Kesharwani(2014)Martin, and Kesharwani]{Martin2014}
Martin,~J. M.~L.; Kesharwani,~M.~K. {Assessment of CCSD(T)-F12 Approximations
  and Basis Sets for Harmonic Vibrational Frequencies}. \emph{J. Chem. Theor.
  Comput.} \textbf{2014}, \emph{10}, 2085--2090\relax
\mciteBstWouldAddEndPuncttrue
\mciteSetBstMidEndSepPunct{\mcitedefaultmidpunct}
{\mcitedefaultendpunct}{\mcitedefaultseppunct}\relax
\EndOfBibitem
\bibitem[Martin \latin{et~al.}(1997)Martin, Taylor, and Lee]{MTLbenzene}
Martin,~J.~M.; Taylor,~P.~R.; Lee,~T.~J. {The harmonic frequencies of benzene.
  A case for atomic natural orbital basis sets}. \emph{Chem. Phys. Lett.}
  \textbf{1997}, \emph{275}, 414--422\relax
\mciteBstWouldAddEndPuncttrue
\mciteSetBstMidEndSepPunct{\mcitedefaultmidpunct}
{\mcitedefaultendpunct}{\mcitedefaultseppunct}\relax
\EndOfBibitem
\bibitem[Werner \latin{et~al.}(2020)Werner, Knowles, Manby, Black, Doll,
  He{\ss}elmann, Kats, K{\"{o}}hn, Korona, Kreplin, Ma, Miller, Mitrushchenkov,
  Peterson, Polyak, Rauhut, and Sibaev]{Werner2020}
Werner,~H.-J. \latin{et~al.}  {The Molpro quantum chemistry package}. \emph{J.
  Chem. Phys.} \textbf{2020}, \emph{152}, 144107\relax
\mciteBstWouldAddEndPuncttrue
\mciteSetBstMidEndSepPunct{\mcitedefaultmidpunct}
{\mcitedefaultendpunct}{\mcitedefaultseppunct}\relax
\EndOfBibitem
\bibitem[Lindh(1993)]{Lindh1993}
Lindh,~R. {The reduced multiplication scheme of the Rys-Gauss quadrature for
  1st order integral derivatives}. \emph{Theor. Chim. Acta} \textbf{1993},
  \emph{85}, 423--440\relax
\mciteBstWouldAddEndPuncttrue
\mciteSetBstMidEndSepPunct{\mcitedefaultmidpunct}
{\mcitedefaultendpunct}{\mcitedefaultseppunct}\relax
\EndOfBibitem
\bibitem[M{\o}ller and Plesset(1934)M{\o}ller, and Plesset]{Moeller1934}
M{\o}ller,~C.; Plesset,~M.~S. {Note on an approximation treatment for
  many-electron systems}. \emph{Phys. Rev.} \textbf{1934}, \emph{46},
  618--622\relax
\mciteBstWouldAddEndPuncttrue
\mciteSetBstMidEndSepPunct{\mcitedefaultmidpunct}
{\mcitedefaultendpunct}{\mcitedefaultseppunct}\relax
\EndOfBibitem
\bibitem[Head-Gordon(1999)]{HeadGordon1999}
Head-Gordon,~M. {An improved semidirect MP2 gradient method}. \emph{Mol. Phys.}
  \textbf{1999}, \emph{96}, 673--679\relax
\mciteBstWouldAddEndPuncttrue
\mciteSetBstMidEndSepPunct{\mcitedefaultmidpunct}
{\mcitedefaultendpunct}{\mcitedefaultseppunct}\relax
\EndOfBibitem
\bibitem[{El-Azhary} \latin{et~al.}(1998){El-Azhary}, Rauhut, Pulay, and
  Werner]{ElAzhary1998}
{El-Azhary},~A.; Rauhut,~G.; Pulay,~P.; Werner,~H.-J. {Analytical energy
  gradients for local second-order M{\o}ller–Plesset perturbation theory}.
  \emph{J. Chem. Phys.} \textbf{1998}, \emph{108}, 5185--5193\relax
\mciteBstWouldAddEndPuncttrue
\mciteSetBstMidEndSepPunct{\mcitedefaultmidpunct}
{\mcitedefaultendpunct}{\mcitedefaultseppunct}\relax
\EndOfBibitem
\bibitem[Watts \latin{et~al.}(1993)Watts, Gauss, and Bartlett]{Watts1993}
Watts,~J.~D.; Gauss,~J.; Bartlett,~R.~J. {Coupled‐cluster methods with
  noniterative triple excitations for restricted open‐shell Hartree–Fock
  and other general single determinant reference functions. Energies and
  analytical gradients}. \emph{J. Chem. Phys.} \textbf{1993}, \emph{98},
  8718--8733\relax
\mciteBstWouldAddEndPuncttrue
\mciteSetBstMidEndSepPunct{\mcitedefaultmidpunct}
{\mcitedefaultendpunct}{\mcitedefaultseppunct}\relax
\EndOfBibitem
\bibitem[Head-Gordon and Head-Gordon(1994)Head-Gordon, and
  Head-Gordon]{HeadGordon1994}
Head-Gordon,~M.; Head-Gordon,~T. {Analytic MP2 frequencies without fifth-order
  storage. Theory and application to bifurcated hydrogen bonds in the water
  hexamer}. \emph{Chem. Phys. Lett.} \textbf{1994}, \emph{220}, 122--128\relax
\mciteBstWouldAddEndPuncttrue
\mciteSetBstMidEndSepPunct{\mcitedefaultmidpunct}
{\mcitedefaultendpunct}{\mcitedefaultseppunct}\relax
\EndOfBibitem
\bibitem[gau()]{gaussian16}
Gaussian 16, Revision C.01, Frisch, M.J., Trucks, G.W., Schlegel, H.B.,
  Scuseria, G.E., Robb, M.A., Cheeseman, J.R.; Scalmani, G.; Barone, V.;
  Petersson, G.A.; Nakatsuji, H.; et al.; Gaussian, Inc., Wallingford, CT
  (2016); http://www.gaussian.com\relax
\mciteBstWouldAddEndPuncttrue
\mciteSetBstMidEndSepPunct{\mcitedefaultmidpunct}
{\mcitedefaultendpunct}{\mcitedefaultseppunct}\relax
\EndOfBibitem
\bibitem[Womack and Manby(2014)Womack, and Manby]{Womack2014}
Womack,~J.~C.; Manby,~F.~R. {Density fitting for three-electron integrals in
  explicitly correlated electronic structure theory}. \emph{J. Chem. Phys.}
  \textbf{2014}, \emph{140}, 044118\relax
\mciteBstWouldAddEndPuncttrue
\mciteSetBstMidEndSepPunct{\mcitedefaultmidpunct}
{\mcitedefaultendpunct}{\mcitedefaultseppunct}\relax
\EndOfBibitem
\bibitem[Gy\"orffy \latin{et~al.}(2017)Gy\"orffy, Knizia, and
  Werner]{Gyorffy2017}
Gy\"orffy,~W.; Knizia,~G.; Werner,~H.-J. {Analytical energy gradients for
  explicitly correlated wave functions. I. Explicitly correlated second-order
  M{\o}ller-Plesset perturbation theory}. \emph{J. Chem. Phys.} \textbf{2017},
  \emph{147}, 214101\relax
\mciteBstWouldAddEndPuncttrue
\mciteSetBstMidEndSepPunct{\mcitedefaultmidpunct}
{\mcitedefaultendpunct}{\mcitedefaultseppunct}\relax
\EndOfBibitem
\bibitem[Gy\"orffy and Werner(2018)Gy\"orffy, and Werner]{Gyorffy2018}
Gy\"orffy,~W.; Werner,~H.-J. {Analytical energy gradients for explicitly
  correlated wave functions. II. Explicitly correlated coupled cluster singles
  and doubles with perturbative triples corrections: CCSD(T)-F12.} \emph{J.
  Chem. Phys.} \textbf{2018}, \emph{148}, 114104\relax
\mciteBstWouldAddEndPuncttrue
\mciteSetBstMidEndSepPunct{\mcitedefaultmidpunct}
{\mcitedefaultendpunct}{\mcitedefaultseppunct}\relax
\EndOfBibitem
\bibitem[Werner \latin{et~al.}(2007)Werner, Adler, and Manby]{Werner2007}
Werner,~H.-J.; Adler,~T.~B.; Manby,~F.~R. {General orbital invariant MP2-F12
  theory}. \emph{J. Chem. Phys.} \textbf{2007}, \emph{126}, 164102\relax
\mciteBstWouldAddEndPuncttrue
\mciteSetBstMidEndSepPunct{\mcitedefaultmidpunct}
{\mcitedefaultendpunct}{\mcitedefaultseppunct}\relax
\EndOfBibitem
\bibitem[Ten-no(2004)]{Ten-no2004}
Ten-no,~S. {Initiation of explicitly correlated Slater-type geminal theory}.
  \emph{Chem. Phys. Lett.} \textbf{2004}, \emph{398}, 56--61\relax
\mciteBstWouldAddEndPuncttrue
\mciteSetBstMidEndSepPunct{\mcitedefaultmidpunct}
{\mcitedefaultendpunct}{\mcitedefaultseppunct}\relax
\EndOfBibitem
\bibitem[Hill \latin{et~al.}(2009)Hill, Peterson, Knizia, and
  Werner]{Hill2009a}
Hill,~J.~G.; Peterson,~K.~A.; Knizia,~G.; Werner,~H.-J. {Extrapolating MP2 and
  CCSD explicitly correlated correlation energies to the complete basis set
  limit with first and second row correlation consistent basis sets}. \emph{J.
  Chem. Phys.} \textbf{2009}, \emph{131}, 194105\relax
\mciteBstWouldAddEndPuncttrue
\mciteSetBstMidEndSepPunct{\mcitedefaultmidpunct}
{\mcitedefaultendpunct}{\mcitedefaultseppunct}\relax
\EndOfBibitem
\bibitem[H{\"{a}}ttig \latin{et~al.}(2010)H{\"{a}}ttig, Tew, and
  K{\"{o}}hn]{Hattig2010}
H{\"{a}}ttig,~C.; Tew,~D.~P.; K{\"{o}}hn,~A. {Communications: Accurate and
  efficient approximations to explicitly correlated coupled-cluster singles and
  doubles, CCSD-F12}. \emph{J. Chem. Phys.} \textbf{2010}, \emph{132},
  231102\relax
\mciteBstWouldAddEndPuncttrue
\mciteSetBstMidEndSepPunct{\mcitedefaultmidpunct}
{\mcitedefaultendpunct}{\mcitedefaultseppunct}\relax
\EndOfBibitem
\bibitem[Weigend and Ahlrichs(2005)Weigend, and Ahlrichs]{Weigend2005}
Weigend,~F.; Ahlrichs,~R. {Balanced basis sets of split valence, triple zeta
  valence and quadruple zeta valence quality for H to Rn: Design and assessment
  of accuracy}. \emph{Phys. Chem. Chem. Phys.} \textbf{2005}, \emph{7},
  3297\relax
\mciteBstWouldAddEndPuncttrue
\mciteSetBstMidEndSepPunct{\mcitedefaultmidpunct}
{\mcitedefaultendpunct}{\mcitedefaultseppunct}\relax
\EndOfBibitem
\bibitem[Rappoport and Furche(2010)Rappoport, and Furche]{Rappoport2010}
Rappoport,~D.; Furche,~F. {Property-optimized Gaussian basis sets for molecular
  response calculations}. \emph{J. Chem. Phys.} \textbf{2010}, \emph{133},
  134105\relax
\mciteBstWouldAddEndPuncttrue
\mciteSetBstMidEndSepPunct{\mcitedefaultmidpunct}
{\mcitedefaultendpunct}{\mcitedefaultseppunct}\relax
\EndOfBibitem
\bibitem[Alml{\"{o}}f and Taylor(1987)Alml{\"{o}}f, and Taylor]{Almlof1987}
Alml{\"{o}}f,~J.; Taylor,~P.~R. {General contraction of Gaussian basis sets. I.
  Atomic natural orbitals for first- and second-row atoms}. \emph{J. Chem.
  Phys.} \textbf{1987}, \emph{86}, 4070--4077\relax
\mciteBstWouldAddEndPuncttrue
\mciteSetBstMidEndSepPunct{\mcitedefaultmidpunct}
{\mcitedefaultendpunct}{\mcitedefaultseppunct}\relax
\EndOfBibitem
\bibitem[Neese and Valeev(2011)Neese, and Valeev]{Neese2011}
Neese,~F.; Valeev,~E.~F. {Revisiting the Atomic Natural Orbital Approach for
  Basis Sets: Robust Systematic Basis Sets for Explicitly Correlated and
  Conventional Correlated {\em ab initio} Methods?} \emph{J. Chem. Theor.
  Comput.} \textbf{2011}, \emph{7}, 33--43\relax
\mciteBstWouldAddEndPuncttrue
\mciteSetBstMidEndSepPunct{\mcitedefaultmidpunct}
{\mcitedefaultendpunct}{\mcitedefaultseppunct}\relax
\EndOfBibitem
\bibitem[Dunning(1989)]{DunningJr1989}
Dunning,~T.~H. {Gaussian basis sets for use in correlated molecular
  calculations. I. The atoms boron through neon and hydrogen}. \emph{J. Chem.
  Phys.} \textbf{1989}, \emph{90}, 1007--1023\relax
\mciteBstWouldAddEndPuncttrue
\mciteSetBstMidEndSepPunct{\mcitedefaultmidpunct}
{\mcitedefaultendpunct}{\mcitedefaultseppunct}\relax
\EndOfBibitem
\bibitem[Kendall \latin{et~al.}(1992)Kendall, Dunning, and
  Harrison]{Kendall1992}
Kendall,~R.~A.; Dunning,~T.~H.; Harrison,~R.~J. {Electron affinities of the
  first‐row atoms revisited. Systematic basis sets and wave functions}.
  \emph{J. Chem. Phys.} \textbf{1992}, \emph{96}, 6796--6806\relax
\mciteBstWouldAddEndPuncttrue
\mciteSetBstMidEndSepPunct{\mcitedefaultmidpunct}
{\mcitedefaultendpunct}{\mcitedefaultseppunct}\relax
\EndOfBibitem
\bibitem[Woon and Dunning(1993)Woon, and Dunning]{Woon1993}
Woon,~D.~E.; Dunning,~T.~H. {Gaussian basis sets for use in correlated
  molecular calculations. III. The atoms aluminum through argon}. \emph{J.
  Chem. Phys.} \textbf{1993}, \emph{98}, 1358--1371\relax
\mciteBstWouldAddEndPuncttrue
\mciteSetBstMidEndSepPunct{\mcitedefaultmidpunct}
{\mcitedefaultendpunct}{\mcitedefaultseppunct}\relax
\EndOfBibitem
\bibitem[Dunning \latin{et~al.}(2001)Dunning, Peterson, and
  Wilson]{Dunning2001}
Dunning,~T.~H.; Peterson,~K.~A.; Wilson,~A.~K. {Gaussian basis sets for use in
  correlated molecular calculations. X. The atoms aluminum through argon
  revisited}. \emph{J. Chem. Phys.} \textbf{2001}, \emph{114}, 9244--9253\relax
\mciteBstWouldAddEndPuncttrue
\mciteSetBstMidEndSepPunct{\mcitedefaultmidpunct}
{\mcitedefaultendpunct}{\mcitedefaultseppunct}\relax
\EndOfBibitem
\bibitem[Martin(1998)]{jmlm107}
Martin,~J. M.~L. {Basis set convergence study of the atomization energy,
  geometry, and anharmonic force field of SO$_2$: The importance of inner
  polarization functions}. \emph{J. Chem. Phys.} \textbf{1998}, \emph{108},
  2791--2800\relax
\mciteBstWouldAddEndPuncttrue
\mciteSetBstMidEndSepPunct{\mcitedefaultmidpunct}
{\mcitedefaultendpunct}{\mcitedefaultseppunct}\relax
\EndOfBibitem
\bibitem[Martin(2006)]{jmlm191}
Martin,~J.~M. {Heats of formation of perchloric acid, HClO$_4$, and perchloric
  anhydride, Cl$_2$O$_7$. Probing the limits of W1 and W2 theory}. \emph{J.
  Mol. Struct. THEOCHEM} \textbf{2006}, \emph{771}, 19--26\relax
\mciteBstWouldAddEndPuncttrue
\mciteSetBstMidEndSepPunct{\mcitedefaultmidpunct}
{\mcitedefaultendpunct}{\mcitedefaultseppunct}\relax
\EndOfBibitem
\bibitem[Peterson and Dunning(2002)Peterson, and Dunning]{Peterson2002}
Peterson,~K.~A.; Dunning,~T.~H. {Accurate correlation consistent basis sets for
  molecular core–valence correlation effects: The second row atoms Al–Ar,
  and the first row atoms B–Ne revisited}. \emph{J. Chem. Phys.}
  \textbf{2002}, \emph{117}, 10548--10560\relax
\mciteBstWouldAddEndPuncttrue
\mciteSetBstMidEndSepPunct{\mcitedefaultmidpunct}
{\mcitedefaultendpunct}{\mcitedefaultseppunct}\relax
\EndOfBibitem
\bibitem[Klopper and Samson(2002)Klopper, and Samson]{Klopper2002a}
Klopper,~W.; Samson,~C. C.~M. {Explicitly correlated second-order
  M{\o}ller–Plesset methods with auxiliary basis sets}. \emph{J. Chem. Phys.}
  \textbf{2002}, \emph{116}, 6397--6410\relax
\mciteBstWouldAddEndPuncttrue
\mciteSetBstMidEndSepPunct{\mcitedefaultmidpunct}
{\mcitedefaultendpunct}{\mcitedefaultseppunct}\relax
\EndOfBibitem
\bibitem[Valeev(2004)]{Valeev2004}
Valeev,~E.~F. {Improving on the resolution of the identity in linear R12 {\em
  ab initio} theories}. \emph{Chem. Phys. Lett.} \textbf{2004}, \emph{395},
  190--195\relax
\mciteBstWouldAddEndPuncttrue
\mciteSetBstMidEndSepPunct{\mcitedefaultmidpunct}
{\mcitedefaultendpunct}{\mcitedefaultseppunct}\relax
\EndOfBibitem
\bibitem[Peterson \latin{et~al.}(2008)Peterson, Adler, and
  Werner]{Peterson2008}
Peterson,~K.~A.; Adler,~T.~B.; Werner,~H.-J. {Systematically convergent basis
  sets for explicitly correlated wavefunctions: The atoms H, He, B–Ne, and
  Al–Ar}. \emph{J. Chem. Phys.} \textbf{2008}, \emph{128}, 084102\relax
\mciteBstWouldAddEndPuncttrue
\mciteSetBstMidEndSepPunct{\mcitedefaultmidpunct}
{\mcitedefaultendpunct}{\mcitedefaultseppunct}\relax
\EndOfBibitem
\bibitem[Yousaf and Peterson(2009)Yousaf, and Peterson]{Yousaf2009}
Yousaf,~K.~E.; Peterson,~K.~A. {Optimized complementary auxiliary basis sets
  for explicitly correlated methods: aug-cc-pVnZ orbital basis sets}.
  \emph{Chem. Phys. Lett.} \textbf{2009}, \emph{476}, 303--307\relax
\mciteBstWouldAddEndPuncttrue
\mciteSetBstMidEndSepPunct{\mcitedefaultmidpunct}
{\mcitedefaultendpunct}{\mcitedefaultseppunct}\relax
\EndOfBibitem
\bibitem[Nielsen(1951)]{Nielsen1951}
Nielsen,~H.~H. {The Vibration-Rotation Energies of Molecules}. \emph{Reviews of
  Modern Physics} \textbf{1951}, \emph{23}, 90--136\relax
\mciteBstWouldAddEndPuncttrue
\mciteSetBstMidEndSepPunct{\mcitedefaultmidpunct}
{\mcitedefaultendpunct}{\mcitedefaultseppunct}\relax
\EndOfBibitem
\bibitem[Marchetti and Werner(2009)Marchetti, and Werner]{Marchetti2009}
Marchetti,~O.; Werner,~H.-J. {Accurate Calculations of Intermolecular
  Interaction Energies Using Explicitly Correlated Coupled Cluster Wave
  Functions and a Dispersion-Weighted MP2 Method}. \emph{J. Phys. Chem. A}
  \textbf{2009}, \emph{113}, 11580--11585\relax
\mciteBstWouldAddEndPuncttrue
\mciteSetBstMidEndSepPunct{\mcitedefaultmidpunct}
{\mcitedefaultendpunct}{\mcitedefaultseppunct}\relax
\EndOfBibitem
\bibitem[Curtiss \latin{et~al.}(1998)Curtiss, Raghavachari, Redfern, Rassolov,
  and Pople]{Curtiss1998a}
Curtiss,~L.~A.; Raghavachari,~K.; Redfern,~P.~C.; Rassolov,~V.; Pople,~J.~A.
  {Gaussian-3 (G3) theory for molecules containing first and second-row atoms}.
  \emph{J. Chem. Phys.} \textbf{1998}, \emph{109}, 7764--7776\relax
\mciteBstWouldAddEndPuncttrue
\mciteSetBstMidEndSepPunct{\mcitedefaultmidpunct}
{\mcitedefaultendpunct}{\mcitedefaultseppunct}\relax
\EndOfBibitem
\bibitem[Mehta \latin{et~al.}(2023)Mehta, Santra, and Martin]{Mehta2023}
Mehta,~N.; Santra,~G.; Martin,~J.~M. {Is explicitly correlated double-hybrid
  density functional theory advantageous for vibrational frequencies?}
  \emph{Canadian Journal of Chemistry} \textbf{2023}, 1--8\relax
\mciteBstWouldAddEndPuncttrue
\mciteSetBstMidEndSepPunct{\mcitedefaultmidpunct}
{\mcitedefaultendpunct}{\mcitedefaultseppunct}\relax
\EndOfBibitem
\bibitem[Liang \latin{et~al.}(2023)Liang, Feng, Liu, and
  Head-Gordon]{Liang2023}
Liang,~J.; Feng,~X.; Liu,~X.; Head-Gordon,~M. {Analytical harmonic vibrational
  frequencies with VV10-containing density functionals: Theory, efficient
  implementation, and benchmark assessments}. \emph{Journal of Chemical
  Physics} \textbf{2023}, \emph{158}, 204109\relax
\mciteBstWouldAddEndPuncttrue
\mciteSetBstMidEndSepPunct{\mcitedefaultmidpunct}
{\mcitedefaultendpunct}{\mcitedefaultseppunct}\relax
\EndOfBibitem
\bibitem[Vydrov and {Van Voorhis}(2010)Vydrov, and {Van Voorhis}]{Vydrov2010}
Vydrov,~O.~A.; {Van Voorhis},~T. {Nonlocal van der Waals density functional:
  The simpler the better}. \emph{The Journal of Chemical Physics}
  \textbf{2010}, \emph{133}, 244103\relax
\mciteBstWouldAddEndPuncttrue
\mciteSetBstMidEndSepPunct{\mcitedefaultmidpunct}
{\mcitedefaultendpunct}{\mcitedefaultseppunct}\relax
\EndOfBibitem
\bibitem[Eckert \latin{et~al.}(1997)Eckert, Pulay, and Werner]{Eckert1997}
Eckert,~F.; Pulay,~P.; Werner,~H.~J. {Ab initio geometry optimization for large
  molecules}. \emph{Journal of Computational Chemistry} \textbf{1997},
  \emph{18}, 1473--1483\relax
\mciteBstWouldAddEndPuncttrue
\mciteSetBstMidEndSepPunct{\mcitedefaultmidpunct}
{\mcitedefaultendpunct}{\mcitedefaultseppunct}\relax
\EndOfBibitem
\bibitem[Smith \latin{et~al.}(2020)Smith, Burns, Simmonett, Parrish, Schieber,
  Galvelis, Kraus, Kruse, {Di Remigio}, Alenaizan, James, Lehtola, Misiewicz,
  Scheurer, Shaw, Schriber, Xie, Glick, Sirianni, O'Brien, Waldrop, Kumar,
  Hohenstein, Pritchard, Brooks, Schaefer, Sokolov, Patkowski, DePrince,
  Bozkaya, King, Evangelista, Turney, Crawford, and Sherrill]{Smith2020}
Smith,~D. G.~A. \latin{et~al.}  {PSI4 1.4: Open-source software for
  high-throughput quantum chemistry}. \emph{The Journal of Chemical Physics}
  \textbf{2020}, \emph{152}, 184108\relax
\mciteBstWouldAddEndPuncttrue
\mciteSetBstMidEndSepPunct{\mcitedefaultmidpunct}
{\mcitedefaultendpunct}{\mcitedefaultseppunct}\relax
\EndOfBibitem
\bibitem[Jensen(2023)]{Jensen2023}
Jensen,~F. {Basis Set Extrapolation of Vibrational Frequencies}. \emph{The
  Journal of Physical Chemistry A} \textbf{2023}, \emph{127}, 2859--2863\relax
\mciteBstWouldAddEndPuncttrue
\mciteSetBstMidEndSepPunct{\mcitedefaultmidpunct}
{\mcitedefaultendpunct}{\mcitedefaultseppunct}\relax
\EndOfBibitem
\bibitem[Varandas(2022)]{Varandas2022}
Varandas,~A. J.~C. {Scale-free-modeling (harmonic) vibrational frequencies:
  Assessing accuracy and cost-effectiveness by CBS extrapolation}. \emph{The
  Journal of Chemical Physics} \textbf{2022}, \emph{157}, 174110\relax
\mciteBstWouldAddEndPuncttrue
\mciteSetBstMidEndSepPunct{\mcitedefaultmidpunct}
{\mcitedefaultendpunct}{\mcitedefaultseppunct}\relax
\EndOfBibitem
\bibitem[Buczek \latin{et~al.}(2011)Buczek, Kupka, and Broda]{Buczek2011a}
Buczek,~A.; Kupka,~T.; Broda,~M.~A. {Estimation of formamide harmonic and
  anharmonic modes in the Kohn-Sham limit using the polarization consistent
  basis sets}. \emph{Journal of Molecular Modeling} \textbf{2011}, \emph{17},
  2265--2274\relax
\mciteBstWouldAddEndPuncttrue
\mciteSetBstMidEndSepPunct{\mcitedefaultmidpunct}
{\mcitedefaultendpunct}{\mcitedefaultseppunct}\relax
\EndOfBibitem
\bibitem[Buczek \latin{et~al.}(2011)Buczek, Kupka, and Broda]{Buczek2011b}
Buczek,~A.; Kupka,~T.; Broda,~M.~A. {Extrapolation of water and formaldehyde
  harmonic and anharmonic frequencies to the B3LYP/CBS limit using polarization
  consistent basis sets}. \emph{Journal of Molecular Modeling} \textbf{2011},
  \emph{17}, 2029--2040\relax
\mciteBstWouldAddEndPuncttrue
\mciteSetBstMidEndSepPunct{\mcitedefaultmidpunct}
{\mcitedefaultendpunct}{\mcitedefaultseppunct}\relax
\EndOfBibitem
\bibitem[Broda \latin{et~al.}(2012)Broda, Buczek, Kupka, and
  Kaminsky]{Broda2012}
Broda,~M.~A.; Buczek,~A.; Kupka,~T.; Kaminsky,~J. {Anharmonic vibrational
  frequency calculations for solvated molecules in the B3LYP Kohn–Sham basis
  set limit}. \emph{Vibrational Spectroscopy} \textbf{2012}, \emph{63},
  432--439\relax
\mciteBstWouldAddEndPuncttrue
\mciteSetBstMidEndSepPunct{\mcitedefaultmidpunct}
{\mcitedefaultendpunct}{\mcitedefaultseppunct}\relax
\EndOfBibitem
\bibitem[Chai and Head-Gordon(2008)Chai, and Head-Gordon]{Chai2008a}
Chai,~J.-D.; Head-Gordon,~M. {Long-range corrected hybrid density functionals
  with damped atom-atom dispersion corrections.} \emph{Physical chemistry
  chemical physics : PCCP} \textbf{2008}, \emph{10}, 6615--6620\relax
\mciteBstWouldAddEndPuncttrue
\mciteSetBstMidEndSepPunct{\mcitedefaultmidpunct}
{\mcitedefaultendpunct}{\mcitedefaultseppunct}\relax
\EndOfBibitem
\bibitem[Jensen(2001)]{Jensen2001}
Jensen,~F. {Polarization consistent basis sets: Principles}. \emph{The Journal
  of Chemical Physics} \textbf{2001}, \emph{115}, 9113--9125\relax
\mciteBstWouldAddEndPuncttrue
\mciteSetBstMidEndSepPunct{\mcitedefaultmidpunct}
{\mcitedefaultendpunct}{\mcitedefaultseppunct}\relax
\EndOfBibitem
\bibitem[Peterson \latin{et~al.}(2015)Peterson, Kesharwani, and
  Martin]{jmlm261}
Peterson,~K.~A.; Kesharwani,~M.~K.; Martin,~J.~M. {The cc-pV5Z-F12 basis set:
  reaching the basis set limit in explicitly correlated calculations}.
  \emph{Mol. Phys.} \textbf{2015}, \emph{113}, 1551--1558\relax
\mciteBstWouldAddEndPuncttrue
\mciteSetBstMidEndSepPunct{\mcitedefaultmidpunct}
{\mcitedefaultendpunct}{\mcitedefaultseppunct}\relax
\EndOfBibitem
\bibitem[Schneider \latin{et~al.}(2008)Schneider, Vogelhuber, Schinle, Stanton,
  and Weber]{Schneider2008}
Schneider,~H.; Vogelhuber,~K.~M.; Schinle,~F.; Stanton,~J.~F.; Weber,~J.~M.
  {Vibrational Spectroscopy of Nitroalkane Chains Using Electron Autodetachment
  and Ar Predissociation}. \emph{J. Phys. Chem. A} \textbf{2008}, \emph{112},
  7498--7506\relax
\mciteBstWouldAddEndPuncttrue
\mciteSetBstMidEndSepPunct{\mcitedefaultmidpunct}
{\mcitedefaultendpunct}{\mcitedefaultseppunct}\relax
\EndOfBibitem
\bibitem[McCaslin and Stanton(2013)McCaslin, and Stanton]{McCaslin2013}
McCaslin,~L.; Stanton,~J. {Calculation of fundamental frequencies for small
  polyatomic molecules: A comparison between correlation consistent and atomic
  natural orbital basis sets}. \emph{Mol. Phys.} \textbf{2013}, \emph{111},
  1492--1496\relax
\mciteBstWouldAddEndPuncttrue
\mciteSetBstMidEndSepPunct{\mcitedefaultmidpunct}
{\mcitedefaultendpunct}{\mcitedefaultseppunct}\relax
\EndOfBibitem
\bibitem[Simandiras \latin{et~al.}(1988)Simandiras, Rice, Lee, Amos, and
  Handy]{Simandiras1988}
Simandiras,~E.~D.; Rice,~J.~E.; Lee,~T.~J.; Amos,~R.~D.; Handy,~N.~C. {On the
  necessity of f basis functions for bending frequencies}. \emph{J. Chem.
  Phys.} \textbf{1988}, \emph{88}, 3187--3195\relax
\mciteBstWouldAddEndPuncttrue
\mciteSetBstMidEndSepPunct{\mcitedefaultmidpunct}
{\mcitedefaultendpunct}{\mcitedefaultseppunct}\relax
\EndOfBibitem
\bibitem[Martin \latin{et~al.}(1998)Martin, Lee, and Taylor]{jmlm104}
Martin,~J. M.~L.; Lee,~T.~J.; Taylor,~P.~R. {A purely {\em ab initio}
  spectroscopic quality quartic force field for acetylene}. \emph{J. Chem.
  Phys.} \textbf{1998}, \emph{108}, 676--691\relax
\mciteBstWouldAddEndPuncttrue
\mciteSetBstMidEndSepPunct{\mcitedefaultmidpunct}
{\mcitedefaultendpunct}{\mcitedefaultseppunct}\relax
\EndOfBibitem
\bibitem[Moran \latin{et~al.}(2006)Moran, Simmonett, Leach, Allen, Schleyer,
  and Schaefer]{Moran2006}
Moran,~D.; Simmonett,~A.~C.; Leach,~F.~E.; Allen,~W.~D.; Schleyer,~P. V.~R.;
  Schaefer,~H.~F. {Popular theoretical methods predict benzene and arenes to be
  nonplanar.} \emph{J. Am. Chem. Soc.} \textbf{2006}, \emph{128},
  9342--9343\relax
\mciteBstWouldAddEndPuncttrue
\mciteSetBstMidEndSepPunct{\mcitedefaultmidpunct}
{\mcitedefaultendpunct}{\mcitedefaultseppunct}\relax
\EndOfBibitem
\bibitem[Nelson \latin{et~al.}(2023)Nelson, Glick, and Sherrill]{Nelson2023}
Nelson,~P.~M.; Glick,~Z.~L.; Sherrill,~C.~D. {Approximating large-basis
  coupled-cluster theory vibrational frequencies using focal-point
  approximations}. \emph{The Journal of chemical physics} \textbf{2023},
  \emph{159}, 094104\relax
\mciteBstWouldAddEndPuncttrue
\mciteSetBstMidEndSepPunct{\mcitedefaultmidpunct}
{\mcitedefaultendpunct}{\mcitedefaultseppunct}\relax
\EndOfBibitem
\bibitem[Kesharwani \latin{et~al.}(2015)Kesharwani, Brauer, and
  Martin]{jmlm260}
Kesharwani,~M.~K.; Brauer,~B.; Martin,~J. M.~L. {Frequency and Zero-Point
  Vibrational Energy Scale Factors for Double-Hybrid Density Functionals (and
  Other Selected Methods): Can Anharmonic Force Fields Be Avoided?} \emph{The
  Journal of Physical Chemistry A} \textbf{2015}, \emph{119}, 1701--1714\relax
\mciteBstWouldAddEndPuncttrue
\mciteSetBstMidEndSepPunct{\mcitedefaultmidpunct}
{\mcitedefaultendpunct}{\mcitedefaultseppunct}\relax
\EndOfBibitem
\bibitem[Laury \latin{et~al.}(2011)Laury, Boesch, Haken, Sinha, Wheeler, and
  Wilson]{Laury2011}
Laury,~M.~L.; Boesch,~S.~E.; Haken,~I.; Sinha,~P.; Wheeler,~R.~A.;
  Wilson,~A.~K. Harmonic vibrational frequencies: Scale factors for pure,
  hybrid, hybrid meta, and double-hybrid functionals in conjunction with
  correlation consistent basis sets. \emph{Journal of Computational Chemistry}
  \textbf{2011}, \emph{32}, 2339--2347\relax
\mciteBstWouldAddEndPuncttrue
\mciteSetBstMidEndSepPunct{\mcitedefaultmidpunct}
{\mcitedefaultendpunct}{\mcitedefaultseppunct}\relax
\EndOfBibitem
\bibitem[Laury \latin{et~al.}(2012)Laury, Carlson, and Wilson]{Laury2012}
Laury,~M.~L.; Carlson,~M.~J.; Wilson,~A.~K. Vibrational frequency scale factors
  for density functional theory and the polarization consistent basis sets.
  \emph{Journal of Computational Chemistry} \textbf{2012}, \emph{33},
  2380--2387\relax
\mciteBstWouldAddEndPuncttrue
\mciteSetBstMidEndSepPunct{\mcitedefaultmidpunct}
{\mcitedefaultendpunct}{\mcitedefaultseppunct}\relax
\EndOfBibitem
\bibitem[{Zapata Trujillo} and McKemmish(2022){Zapata Trujillo}, and
  McKemmish]{ZapataTrujillo2022}
{Zapata Trujillo},~J.~C.; McKemmish,~L.~K. {VIBFREQ1295: A New Database for
  Vibrational Frequency Calculations}. \emph{The Journal of Physical Chemistry
  A} \textbf{2022}, \emph{126}, 4100--4122\relax
\mciteBstWouldAddEndPuncttrue
\mciteSetBstMidEndSepPunct{\mcitedefaultmidpunct}
{\mcitedefaultendpunct}{\mcitedefaultseppunct}\relax
\EndOfBibitem
\bibitem[{Zapata Trujillo} and McKemmish(2023){Zapata Trujillo}, and
  McKemmish]{ZapataTrujillo2023}
{Zapata Trujillo},~J.~C.; McKemmish,~L.~K. {Model Chemistry Recommendations for
  Scaled Harmonic Frequency Calculations: A Benchmark Study}. \emph{The Journal
  of Physical Chemistry A} \textbf{2023}, \emph{127}, 1715--1735\relax
\mciteBstWouldAddEndPuncttrue
\mciteSetBstMidEndSepPunct{\mcitedefaultmidpunct}
{\mcitedefaultendpunct}{\mcitedefaultseppunct}\relax
\EndOfBibitem
\bibitem[Schneider and Thiel(1989)Schneider, and Thiel]{Schneider1989}
Schneider,~W.; Thiel,~W. {Anharmonic force fields from analytic second
  derivatives: method and application to methyl bromide}. \emph{Chem. Phys.
  Lett.} \textbf{1989}, \emph{157}, 367--373\relax
\mciteBstWouldAddEndPuncttrue
\mciteSetBstMidEndSepPunct{\mcitedefaultmidpunct}
{\mcitedefaultendpunct}{\mcitedefaultseppunct}\relax
\EndOfBibitem
\bibitem[Morgante and Peverati(2019)Morgante, and Peverati]{Morgante2019}
Morgante,~P.; Peverati,~R. {ACCDB: A collection of chemistry databases for
  broad computational purposes}. \emph{Journal of Computational Chemistry}
  \textbf{2019}, \emph{40}, 839--848\relax
\mciteBstWouldAddEndPuncttrue
\mciteSetBstMidEndSepPunct{\mcitedefaultmidpunct}
{\mcitedefaultendpunct}{\mcitedefaultseppunct}\relax
\EndOfBibitem
\bibitem[Kozuch and Martin(2013)Kozuch, and Martin]{Kozuch2013}
Kozuch,~S.; Martin,~J. M.~L. {Spin-component-scaled double hybrids: An
  extensive search for the best fifth-rung functionals blending DFT and
  perturbation theory}. \emph{Journal of Computational Chemistry}
  \textbf{2013}, \emph{34}, 2327–2344\relax
\mciteBstWouldAddEndPuncttrue
\mciteSetBstMidEndSepPunct{\mcitedefaultmidpunct}
{\mcitedefaultendpunct}{\mcitedefaultseppunct}\relax
\EndOfBibitem
\bibitem[Santra \latin{et~al.}(2019)Santra, Sylvetsky, and Martin]{Santra2019a}
Santra,~G.; Sylvetsky,~N.; Martin,~J. M.~L. {Minimally Empirical Double-Hybrid
  Functionals Trained against the GMTKN55 Database: revDSD-PBEP86-D4,
  revDOD-PBE-D4, and DOD-SCAN-D4}. \emph{The Journal of Physical Chemistry A}
  \textbf{2019}, \emph{123}, 5129--5143\relax
\mciteBstWouldAddEndPuncttrue
\mciteSetBstMidEndSepPunct{\mcitedefaultmidpunct}
{\mcitedefaultendpunct}{\mcitedefaultseppunct}\relax
\EndOfBibitem
\bibitem[Sylvetsky and Martin(2019)Sylvetsky, and Martin]{jmlm281}
Sylvetsky,~N.; Martin,~J. M.~L. {Probing the basis set limit for thermochemical
  contributions of inner-shell correlation: balance of core-core and
  core-valence contributions}. \emph{Mol. Phys.} \textbf{2019}, \emph{117},
  1078--1087\relax
\mciteBstWouldAddEndPuncttrue
\mciteSetBstMidEndSepPunct{\mcitedefaultmidpunct}
{\mcitedefaultendpunct}{\mcitedefaultseppunct}\relax
\EndOfBibitem
\bibitem[Dunham(1932)]{Dunham1932}
Dunham,~J.~L. The Wentzel-Brillouin-Kramers Method of Solving the Wave
  Equation. \emph{Phys. Rev.} \textbf{1932}, \emph{41}, 713--720\relax
\mciteBstWouldAddEndPuncttrue
\mciteSetBstMidEndSepPunct{\mcitedefaultmidpunct}
{\mcitedefaultendpunct}{\mcitedefaultseppunct}\relax
\EndOfBibitem
\bibitem[Gauss \latin{et~al.}(2006)Gauss, Tajti, Kállay, Stanton, and
  Szalay]{DBOC_CCSD}
Gauss,~J.; Tajti,~A.; Kállay,~M.; Stanton,~J.~F.; Szalay,~P.~G. {Analytic
  calculation of the diagonal Born-Oppenheimer correction within
  configuration-interaction and coupled-cluster theory}. \emph{The Journal of
  Chemical Physics} \textbf{2006}, \emph{125}, 144111\relax
\mciteBstWouldAddEndPuncttrue
\mciteSetBstMidEndSepPunct{\mcitedefaultmidpunct}
{\mcitedefaultendpunct}{\mcitedefaultseppunct}\relax
\EndOfBibitem
\bibitem[Matthews \latin{et~al.}(2020)Matthews, Cheng, Harding, Lipparini,
  Stopkowicz, Jagau, Szalay, Gauss, and Stanton]{CFOUR}
Matthews,~D.~A.; Cheng,~L.; Harding,~M.~E.; Lipparini,~F.; Stopkowicz,~S.;
  Jagau,~T.-C.; Szalay,~P.~G.; Gauss,~J.; Stanton,~J.~F. {Coupled-cluster
  techniques for computational chemistry: The CFOUR program package}. \emph{The
  Journal of Chemical Physics} \textbf{2020}, \emph{152}, 214108\relax
\mciteBstWouldAddEndPuncttrue
\mciteSetBstMidEndSepPunct{\mcitedefaultmidpunct}
{\mcitedefaultendpunct}{\mcitedefaultseppunct}\relax
\EndOfBibitem
\bibitem[Miani \latin{et~al.}(2000)Miani, Can{\'{e}}, Palmieri, Trombetti, and
  Handy]{Miani2000}
Miani,~A.; Can{\'{e}},~E.; Palmieri,~P.; Trombetti,~A.; Handy,~N.~C.
  {Experimental and theoretical anharmonicity for benzene using density
  functional theory}. \emph{J. Chem. Phys.} \textbf{2000}, \emph{112},
  248--259\relax
\mciteBstWouldAddEndPuncttrue
\mciteSetBstMidEndSepPunct{\mcitedefaultmidpunct}
{\mcitedefaultendpunct}{\mcitedefaultseppunct}\relax
\EndOfBibitem
\bibitem[Demaison(2007)]{Demaison2007}
Demaison,~J. {Experimental, semi-experimental and {\em ab initio} equilibrium
  structures}. \emph{Mol. Phys.} \textbf{2007}, \emph{105}, 3109--3138\relax
\mciteBstWouldAddEndPuncttrue
\mciteSetBstMidEndSepPunct{\mcitedefaultmidpunct}
{\mcitedefaultendpunct}{\mcitedefaultseppunct}\relax
\EndOfBibitem
\bibitem[Eriksen \latin{et~al.}(2020)Eriksen, Anderson, Deustua, Ghanem, Hait,
  Hoffmann, Lee, Levine, Magoulas, Shen, Tubman, Whaley, Xu, Yao, Zhang, Alavi,
  Chan, Head-Gordon, Liu, Piecuch, Sharma, Ten-no, Umrigar, and
  Gauss]{Eriksen2020}
Eriksen,~J.~J. \latin{et~al.}  {The Ground State Electronic Energy of Benzene}.
  \emph{J. Phys. Chem. Lett.} \textbf{2020}, \emph{11}, 8922--8929\relax
\mciteBstWouldAddEndPuncttrue
\mciteSetBstMidEndSepPunct{\mcitedefaultmidpunct}
{\mcitedefaultendpunct}{\mcitedefaultseppunct}\relax
\EndOfBibitem
\bibitem[Tao \latin{et~al.}(2003)Tao, Perdew, Staroverov, and
  Scuseria]{Tao2003}
Tao,~J.; Perdew,~J.~P.; Staroverov,~V.~N.; Scuseria,~G.~E. {Climbing the
  Density Functional Ladder: Nonempirical Meta–Generalized Gradient
  Approximation Designed for Molecules and Solids}. \emph{Physical Review
  Letters} \textbf{2003}, \emph{91}, 146401\relax
\mciteBstWouldAddEndPuncttrue
\mciteSetBstMidEndSepPunct{\mcitedefaultmidpunct}
{\mcitedefaultendpunct}{\mcitedefaultseppunct}\relax
\EndOfBibitem
\bibitem[Lee \latin{et~al.}(1988)Lee, Yang, and Parr]{Lee1988a}
Lee,~C.; Yang,~W.; Parr,~R.~G. {Development of the Colle-Salvetti
  correlation-energy formula into a functional of the electron density}.
  \emph{Physical Review B} \textbf{1988}, \emph{37}, 785--789\relax
\mciteBstWouldAddEndPuncttrue
\mciteSetBstMidEndSepPunct{\mcitedefaultmidpunct}
{\mcitedefaultendpunct}{\mcitedefaultseppunct}\relax
\EndOfBibitem
\bibitem[Becke(1993)]{Becke1993a}
Becke,~A.~D. {A new mixing of Hartree–Fock and local density‐functional
  theories}. \emph{The Journal of Chemical Physics} \textbf{1993}, \emph{98},
  1372--1377\relax
\mciteBstWouldAddEndPuncttrue
\mciteSetBstMidEndSepPunct{\mcitedefaultmidpunct}
{\mcitedefaultendpunct}{\mcitedefaultseppunct}\relax
\EndOfBibitem
\bibitem[Lin \latin{et~al.}(2013)Lin, Li, Mao, and Chai]{Lin2013}
Lin,~Y.-S.; Li,~G.-D.; Mao,~S.-P.; Chai,~J.-D. {Long-Range Corrected Hybrid
  Density Functionals with Improved Dispersion Corrections}. \emph{Journal of
  Chemical Theory and Computation} \textbf{2013}, \emph{9}, 263--272\relax
\mciteBstWouldAddEndPuncttrue
\mciteSetBstMidEndSepPunct{\mcitedefaultmidpunct}
{\mcitedefaultendpunct}{\mcitedefaultseppunct}\relax
\EndOfBibitem
\bibitem[Karton \latin{et~al.}(2008)Karton, Tarnopolsky, Lam{\`{e}}re, Schatz,
  and Martin]{Karton2008}
Karton,~A.; Tarnopolsky,~A.; Lam{\`{e}}re,~J.-F.; Schatz,~G.~C.; Martin,~J.
  M.~L. {Highly Accurate First-Principles Benchmark Data Sets for the
  Parametrization and Validation of Density Functional and Other Approximate
  Methods. Derivation of a Robust, Generally Applicable, Double-Hybrid
  Functional for Thermochemistry and Thermochemical }. \emph{The Journal of
  Physical Chemistry A} \textbf{2008}, \emph{112}, 12868--12886\relax
\mciteBstWouldAddEndPuncttrue
\mciteSetBstMidEndSepPunct{\mcitedefaultmidpunct}
{\mcitedefaultendpunct}{\mcitedefaultseppunct}\relax
\EndOfBibitem
\bibitem[Mardirossian and Head-Gordon(2014)Mardirossian, and
  Head-Gordon]{Mardirossian2014}
Mardirossian,~N.; Head-Gordon,~M. {$\omega$B97X-V: A 10-parameter,
  range-separated hybrid, generalized gradient approximation density functional
  with nonlocal correlation, designed by a survival-of-the-fittest strategy}.
  \emph{Physical Chemistry Chemical Physics} \textbf{2014}, \emph{16},
  9904\relax
\mciteBstWouldAddEndPuncttrue
\mciteSetBstMidEndSepPunct{\mcitedefaultmidpunct}
{\mcitedefaultendpunct}{\mcitedefaultseppunct}\relax
\EndOfBibitem
\bibitem[Mardirossian and Head-Gordon(2016)Mardirossian, and
  Head-Gordon]{Mardirossian2016}
Mardirossian,~N.; Head-Gordon,~M. {$\omega$B97M-V: A combinatorially optimized,
  range-separated hybrid, meta-GGA density functional with VV10 nonlocal
  correlation}. \emph{The Journal of Chemical Physics} \textbf{2016},
  \emph{144}, 214110\relax
\mciteBstWouldAddEndPuncttrue
\mciteSetBstMidEndSepPunct{\mcitedefaultmidpunct}
{\mcitedefaultendpunct}{\mcitedefaultseppunct}\relax
\EndOfBibitem
\bibitem[Mardirossian and Head-Gordon(2015)Mardirossian, and
  Head-Gordon]{Mardirossian2015}
Mardirossian,~N.; Head-Gordon,~M. {Mapping the genome of meta-generalized
  gradient approximation density functionals: The search for B97M-V}. \emph{The
  Journal of Chemical Physics} \textbf{2015}, \emph{142}, 074111\relax
\mciteBstWouldAddEndPuncttrue
\mciteSetBstMidEndSepPunct{\mcitedefaultmidpunct}
{\mcitedefaultendpunct}{\mcitedefaultseppunct}\relax
\EndOfBibitem
\end{mcitethebibliography}
\end{document}